\documentclass[useAMS,usenatbib]{mn2e}

\usepackage{aas_macros}
\usepackage{amsmath}
\usepackage{amssymb}
\usepackage{float}
\usepackage{graphicx}
\usepackage{lscape}
\usepackage{url}
\usepackage{natbib}
\usepackage{placeins}
\usepackage{color}
\usepackage{wasysym}
\usepackage{hyperref}
\usepackage{scrextend}
\DeclareMathOperator\erf{erf}
\newcommand{\nhii}{\ensuremath{n_{{\mbox{\small{H}}}_{\mbox{\small{II}}}}}}
\newcommand{\nhi}{\ensuremath{n_{\mbox{\small{H}}}}}

\begin{document}

\title{ Hot planetary winds near a star: dynamics,    wind-wind interactions, and observational signatures}

\author[J. Carroll-Nellenback, A. Frank, B.Liu, A.C. Quillen, E.G. Blackman\&  I. Dobbs-Dixon]
{
Jonathan Carroll-Nellenback$^{1}$\thanks{E-mail:jonathan.carroll@rochester.edu}, Adam Frank$^{2}$, Baowei Liu$^{1,2}$,\newauthor  Alice C. Quillen$^{2}$, Eric G. Blackman$^{2}$ and Ian Dobbs-Dixon$^{3}$\\
$^{1}$Center for Integrated Research Computing, University of Rochester, Rochester NY 14627\\
$^{2}$Department of Physics and Astronomy, University of Rochester, Rochester NY 14627\\
$^{3}$New York University Abu Dhabi, Abu Dhabi, United Arab Emirates
}

\date{Submitted 2015 April 27}

\pagerange{\pageref{firstpage}--\pageref{lastpage}}
\maketitle
\label{firstpage}

\begin{abstract}
Signatures of ``evaporative'' winds from exo-planets on short (hot) orbits around their host star have been observed in a number of systems.  In this paper we present global AMR simulations that track the launching of the winds, their expansion through the circumstellar environment, and their interaction with a stellar wind.  We focus on purely hydrodynamic flows including the anisotropy of the wind launching and explore the orbital/fluid dynamics of the resulting flows in detail.  In particular we  find that a combination of the tidal and Coriolis forces strongly distorts the planetary "Parker" wind creating ``up-orbit'' and ``down-orbit'' streams. We characterize the flows in terms of their orbital elements which change depending on their launch position on the planet.  We find that the anisotropy in the atmospheric temperature leads to significant backflow on to the planet.  The planetary wind interacts strongly with the stellar wind creating instabilities that cause eventual deposition of planetary gas onto the star.  We present synthetic observations of both transit and absorption line-structure for our simulations. For our initial conditions, we find that  the orbiting wind material produces absorption signatures at significant distances from the planet and substantial orbit to orbit variability.  Ly-$\alpha$ absorption shows red and blueshifted features out to $70$ km/s.  Finally, using semi-analytic models we constrain the effect of radiation pressure, given the approximation of uniform stellar absorption.

\end{abstract}


\section{Introduction}

Understanding planetary atmospheres 
has been an emerging pillar of  exoplanet science.  
These atmospheres and their interactions 
with their near-space environment
provide a fertile opportunity for 
comparing observations with theoretical predictions.  In particular, atmospheric “blow-off”, also known as hydrodynamic escape or evaporation, has a strong influence on  key facets of planetary evolution from end-state masses to final atmospheric compositions--and therefore habitability.  For some exoplanets (i.e. HD 209458b, \cite{ballister07} ), the observational signatures of atmospheric blow-off are determined and  provide a rich source of diagnostics for theoretical models.  

The implications of  understanding planetary winds are far reaching. Atmospheric escape has, for example, played an important role in shaping the early evolution of  planetary atmospheres in the Solar system (e.g., \cite{zahnle86,hunten87}). Evaporative winds and atmospheric escape from exoplanets may  be critical (e.g. \cite{tarter07,seager09}) in determining the habitability of super-Earths and Earth-like planets. 

For close-in planets, a significant fraction of a planet's initial mass can be evaporated due to strong EUV radiation from a young star \citep{lopez14,perezbecker13,OwenWu2013,OwenAdams2016}.
Atmospheric circulation and blow-off are  strongly coupled to planetary dynamics  since rotation can induce asymmetry and structure in the planetary wind, but can also cause the planet's orbit to drift via an analogy to the Yarkovsky effect \citep{teyssandier15}.

Observationally, the problem of atmospheric blow off has already yielded  a handful of high quality datasets. The extended atmosphere and/or wind from close-in transiting exoplanets have been probed by transmission spectroscopy using strong atomic resonance lines, predominately at UV wavelengths.  Transit observations in the Lyman-α line of neutral hydrogen gave the first detection of atmospheric escape from an exoplanet, the hot-Jupiter HD 209458b \citep{vidalmadjar03}. The extended cloud of neutral hydrogen surrounding the planet was confirmed by subsequent observations with STIS and The Cosmic Origins Spectrograph (COS) \citep{vidalmadjar04,vidalmadjar13,ehrenreich08,benjaffel10}.

Heavier elements such as MgI, MgII, OI, CII, and SIII have also been detected at high altitudes from HD 209458b and WASP-12b \citep{vidalmadjar04,benjaffel10,linsky10,vidalmadjar13,fossati10} suggesting these atmospheres are in a hydrodynamic blow-off  state as well. Multi-epoch spectra exhibit variations in the Ly-$\alpha$ transit depth, which are correlated with flares from the host star, have been detected with HST and SWIFT \citep{lecavelier12}. 

Many open questions about  these winds  require improved  physical modeling.  For example the inferred wide distribution of neutral atom velocities in exoplanet thermosphere’s has been difficult to explain (e.g.,\cite{Koskinenea2013}), and estimated mass outflow rates are uncertain because they depend on assumed isotropy.  Furthermore, physical processes that could affect the wind structure and outflow rates, such as planetary magnetic fields \citep{OwenAdams2014}, time dependent EUV flux \citep{lecavelier12}, atmospheric circulation \citep{teyssandier15} and the interaction between stellar and planet winds \citep{murrayclay09,stone09}  have only recently begun to be incorporated into existing simulations. Until recently, only a few groups have  carried out fully 3-D simulations \citep{schneiter07,cohen11,bisikalo13}. More recent studies have begun exploring global dynamics at higher resolution with more physics. Matsakos et al 2015, for example, presents a 3-D MHD study that tracked the full orbital dynamics of the wind (see also \citep{schneiter16,tripathi15,christie16}) 

In this paper we present initial studies of the interaction between  planetary  and stellar winds.  Using Adaptive Mesh Refinement (AMR) simulations we explore the hydrodynamic evolution of an anisotropic wind launched from a hot Jupiter in orbit, as it interacts with the stellar wind.  We also create synthetic observations of our system that includes $Ly_\alpha$ line profiles and transit light curves. This paper is intended to be the first  of a  long term study using AMR methodologies to investigate atmospheric blow-off in its global (i.e full orbital) context. 

Our paper is structured as follows:  Section II provides details of our method and model. Section III presents and discusses our simulation results.  In section IV we present our synthetic observations and interpret them.  In section V we discuss the limitations of our simulations and open questions for future work.  We conclude in section VI.

\section{Methods and Model}
Simulations were conducted using AstroBEAR \footnote{https://astrobear.pas.rochester.edu/} \citep{cunningham09,carroll13}, a publicly available, massively parallelized, adaptive mesh refinement (AMR) code that contains a variety of multiphysics solvers (i.e. self-gravity, magnetic resistivity, radiative transport, ionization dynamics, heat conduction, etc).  The models in this paper emerge from  solving  the equations of fluid dynamics in a co-rotating frame, including the gravitational force from both the star and the planet, and  tracking the ionization state of the gas. The mass, momentum and energy equations solved are
\begin{equation}
    \frac{\partial \rho}{\partial t} + \boldsymbol{\nabla} \cdot \rho \boldsymbol{v} = 0 
    \label{eq:Eu1}
\end{equation}
\begin{equation}
    \frac{\partial \rho \boldsymbol{v}}{\partial t} + \boldsymbol{\nabla} \cdot \left ( \rho \boldsymbol{v} \otimes \boldsymbol{v} \right )= - \nabla P - \rho \nabla \phi + \boldsymbol{f_R}
    \label{eq:Eu2}
\end{equation}
\begin{equation}
    \frac{\partial E}{\partial t} + \boldsymbol{\nabla} \cdot ((E + P) \boldsymbol{v}) = 0
    \label{eq:Eu3}
\end{equation}
where $\rho$ is the mass density, $\boldsymbol{v}$ is the velocity, $P$ is the thermal pressure, and the total energy $E = \frac{1}{\gamma - 1} P + \frac{1}{2}\rho v^2$.  We set $\gamma = 1.0001$ to keep the planetary and stellar gas at constant temperatures.  
The gravitational potential $\phi$ includes the gravity of both the star and the planet, and the Coriolis and centrifugal forces are given by $\boldsymbol{f_R} = \rho \left ( - 2 \boldsymbol{\Omega} \times \boldsymbol{v} - \boldsymbol{\Omega} \times \left ( \boldsymbol{\Omega} \times \boldsymbol{r} \right ) \right )$

The simulations also tracked the advection and the ionization/recombination of neutral ($n_{\mbox{\small{H}}}$) and ionized (\nhii) hydrogen species.
\begin{equation}
    \frac{\partial \nhi}{\partial t} + \boldsymbol{\nabla} \cdot \nhi \boldsymbol{v} = -\Gamma
    \label{eq:Eu4}
\end{equation}
\begin{equation}
    \frac{\partial \nhii}{\partial t} + \boldsymbol{\nabla} \cdot \nhii \boldsymbol{v} = \Gamma
    \label{eq:Eu5}
\end{equation}
where the rate of ionization per unit volume
is given by
\begin{equation}
\Gamma = \nhi n_{\mbox{\small{e}}} \Gamma_c - \nhii n_{\mbox{\small{e}}} \Gamma_r + \nhi A_{\star}\left (\frac{r}{\mbox{AU}} \right )^{-2}
\label{eq:ion}
\end{equation}
where the three terms on the right correspond to   collisional ionization \citep{arnaud85}, radiative recombination \citep{verner96}, and photo-ionization respectively. $A_\star$ in Equation \ref{eq:ion} is the ionization rate at $1 \mbox{AU}$.  The model also assumes $n_{\mbox{\small{e}}} = \nhii$.  Note because of the isothermal equation of state, the ionization state of the gas did not alter the pressure or the dynamics of the system.

\subsection{Model System}

The simulated binary system is modelled after HD204958b, a $0.69 M_{\jupiter}$ planet with a radius of $1.38 R_{\jupiter}$ orbiting at a distance of $a=.04747 \mbox{ AU}$ around a $1.148 M_{\sun}$ star with a radius of $1.203 R_{\sun}$ \footnote{http://exoplanet.eu}.  An ionizing flux $ A_{\star} = 8\times 10^{-9} \mbox{s}^{-1}$ was chosen which gives an ionization time scale $\tau_H=\left ( \frac{A_{\star}}{0.04747^2} \right ) ^{-1} = 3.26 \mbox{ days}$ at the distance of HD204958b.
This is comparable to the orbital period $P=3.524 \mbox{ days}$ and 1/10th that of the Sun's ionizing flux at solar minimum \citep{bzowski08}. 

Table \ref{tab:parameters} lists the parameters used in the simulations.
\setcounter{table}{0}

\begin{table}
\begin{minipage}{80mm}
\renewcommand{\thempfootnote}{\fnsymbol{footnote}}

\caption{Parameters used in the simulations}
\label{tab:parameters}
\begin{tabular}{l|c|c|}\hline

Planet Mass &$M_{p}$ & $0.69 M_{\jupiter}$ \footnote{\label{exoplanet} Exoplanet Encyclopedia (http://exoplanet.eu)} \\
Planet Radius &$R_{p}$ & $1.38 R_{\jupiter}$ \footref{exoplanet} \\
Planet Temperature &$T_{p}$ & $10^{4} K$ \\
Planet Escape Parameter &$\lambda_{p}$ & $10.738$ \\
Planet Density &$\rho_{p}$  & $3.21\times 10^{-15} \mbox{g cm}^{-3}$ \\
Stellar Mass &$M_{\star}$ & $1.148 M_{\sun}$ \footref{exoplanet} \\
Stellar Radius &$R_{\star}$ & $1.203 R_{\sun}$ \footref{exoplanet} \\
Stellar Temperature &$T_{\star}$ & $10^{6} K$ \\
Stellar Escape Parameter &$\lambda_{\star}$ & $22.088$ \\
Stellar Density &$\rho_{*}$  & $3.21 \times 10^{-13} \mbox{g cm}^{-3}$ \\
Mass Ratio (planet/star) & $q$ & $5.74\times 10^{-4}$ \\  
Orbital Separation & $a$ & $0.04747$ AU \footref{exoplanet} \\
Planet Orbital Radius & $r_p$ & $0.04744$ AU \\
Orbital Period & $P$ & $3.524$ days \\
Ionization Timescale & $\tau_H$ & $3.26$ days \\
 \hline
\end{tabular}
\end{minipage}
\end{table}

\subsection{Planetary winds}

Planetary blow-off occurs when irradiation from the central star, especially in the extreme ultraviolet (EUV), heats of the upper layers of the atmosphere to produce an extended envelope of gas which transitions into an wind.  The parameter $\lambda$ (to appear later in Equation \ref{eq:parker}) is called the hydrodynamic escape parameter and is the ratio of gravitational potential to thermal energy at the top of the atmosphere.  It is also a characteristic measure of the strength of the wind.  For $\lambda \gg 10$, the atmosphere is too tightly bound for a hydrodynamic wind to form.  Note that weaker outflows may be produced via non-thermal processes (e.g., \cite{hunten82}). For $\lambda \sim 10$, a Parker-type thermally driven hydrodynamic wind is expected (note that $\lambda  \sim 15$ for the sun with its $T \sim 10^6$ K corona). Note that extended subsonic regimes are possible for larger $\lambda$ though the production of a wind that escapes to infinity will require the transition to a supersonic flow with $v_w >v_{esc}$

The actual temperature of a planet’s thermosphere (determined by its composition and stellar radiative energy input) is a source of ongoing debate and it's here that a proper accounting of the physics of stellar energy deposition in the atmosphere is needed.   When evaluated at the effective temperature $T_{\mbox{eff}}\approx 10^3 \mbox{K}$, the exoplanet HD209458b yields $\lambda \approx 140$, indicating that a hydrodynamic wind is unlikely. However, EUV radiation from the central star can heat the upper layers of planets on “hot” orbits to $T \sim 10^4 \mbox{K}$ \citep{lammer03,yelle04,ballister07}. If the gas is neutral the escape parameter becomes $\lambda \approx 14$ (akin to that in the solar corona). Thus thermally driven winds are possible - and to be expected -  for planets on hot orbits.  Partially ionized, or higher temperature atmospheres will yield even smaller values of $\lambda$, (partial ionization which is expected based on theoretical models \citep{murrayclay09,ownalvarez}).  This is particularly relevant for planets orbiting young stars with their higher levels of activity.

\subsection{Isothermal Winds}
We input the winds into our simulations as  outflows from a spherical boundary with a fixed temperature, velocity, and density. Given our choice of $\gamma$ we expect isothermal winds (i.e. Parker winds) to be generated.  In this section we discuss the analytic isothermal wind solution that was used to initialize our models.

The \cite{parker58}
spherical isothermal wind solution from a star with mass $M$, radius $R$,  and sound speed $c_s$ is given by 
\begin{equation}
    \psi - \log \psi = -3 -4 \log \frac{\lambda}{2} + 4 \log \xi + \frac{2 \lambda}{\xi}
  \label{eq:parker}
\end{equation}
where $\psi = \frac{v^2}{c_s^2}$, $\xi = \frac{r}{R}$, and $\lambda=\frac{G M \mu }{R k_B T} $ where $\mu$ is the mean mass per particle.

We can eliminate the dependence of the solution on $\lambda$ by scaling $r$ 
in units of the sonic radius  $r_s = \frac{\lambda R}{2}$, where the flow transitions from a subsonic atmosphere to a supersonic wind.  Defining $\Xi = \frac{2\xi}{\lambda} = \frac{r}{r_s}$ Equation \ref{eq:parker} becomes
\begin{equation}
      \psi - \log \psi - 1 = 4 \left ( \Xi^{-1} - \log \Xi^{-1} - 1 \right ),
\end{equation}
where the star or planet boundary is located at $\Xi_0 = \frac{2}{\lambda}$.  Note that there are $\frac{\lambda}{2}$ stellar radii until the solution reaches the sonic point, so stars (or planets) with large $\lambda$ will have extended subsonic atmospheres, while stars or planets with a small $\lambda$ will quickly transition to a supersonic wind.

Expressing the density as $\phi = \frac{\rho}{\rho_s}$ where $\rho_s$ is the density at the sonic point, and using the conservation of mass, we have
\begin{equation}
  \phi \Xi^2 \psi^{1/2} = 1,
\end{equation}
which combined with Equation \ref{eq:parker} gives
\begin{equation}
      \phi = \exp{\frac{2}{\Xi} - \frac{3}{2} - \frac{\psi}{2}}.
\end{equation}

\subsection{Description of simulations}

\begin{figure}
\centering
\includegraphics[width=\columnwidth]{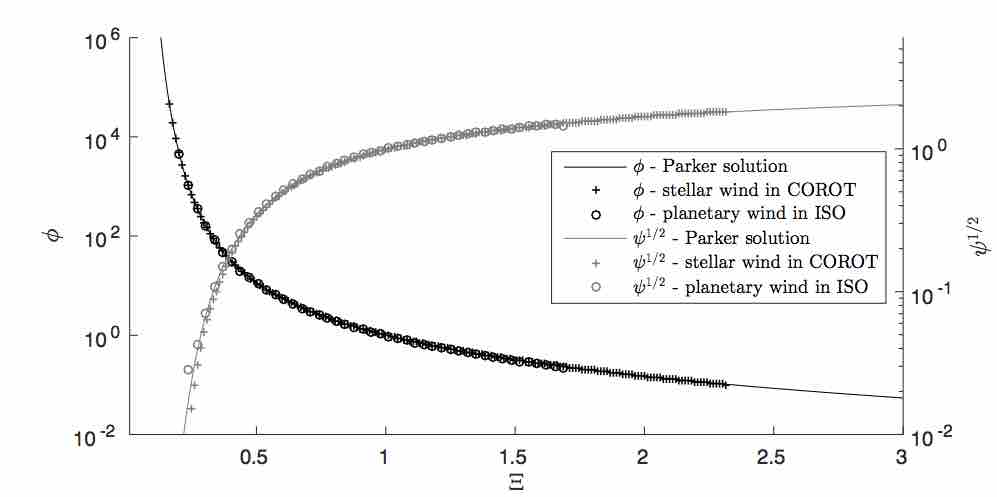}
\caption{Plot of the isothermal Parker wind solution.  $\psi^{1/2}$ is the mach number of the wind, $\phi$ represents the density normalized to the density at the sonic point $\rho_s$, and $\Xi$ is the radius normalized to the sonic radius $r_s$.  Also shown are the steady state solutions for the planetary wind used in run ISO, and the stellar wind used in run ROT. }
\label{fig:parkerwind2}
\end{figure}

Our simplest model, labeled ISO, consists of an isotropic wind emanating from a planet modeled after HD209458b with a temperature of $T_p=10^4 K$ and a corresponding $\lambda_p=10.738$.  The density at the planet surface $\rho_p=3.2112 \times 10^{-15} \mbox{ g cm}^{-3}$ was chosen to give a theoretical mass loss rate of $\dot{M}_p=10^{10} \mbox {g s}^{-1}$.  Our planetary model differs slightly from an exact Parker wind in that the velocity at the surface is set to $0$ instead of the analytic value of $c_s \psi^{1/2}$, however this does not alter the solution significantly. We initialize the grid with the Parker solution, and then hold the density and velocity fixed at the planet surface and allow the simulation to relax.  Figure \ref{fig:parkerwind2} shows the angle averaged density and radial velocity scaled to the dimensionless parameters $\phi$ and $\psi$.

Our second model, labeled ANISO, uses an anisotropic planetary boundary with a density and temperature profile similar to \cite{stone09}.  The day side of the planet has a temperature $T(\theta)=T_{p} \max \left [ 0.01, \cos (\theta) \right ]$ where $\theta$ is the angle of incidence of the light from the star (and the angular distance from the substellar point).  
The night-side is kept at $.01 T_p = 100 \mbox{K}$.  We initialize the simulation with an isotropic Parker wind as in ISO, and then allow the simulation to relax to the anisotropic planetary boundary.  As in \cite {stone09} we expect the temperature variation to lead to strong winds from the day side towards the night side. 

Our third and fourth models, labeled ISOROT and ANISOROT have the same setup as ISO and ANISO respectively, but now include the gravitational force from a stellar companion as well as the Coriolis and centripetal forces in the co-rotating frame.  We also modify the velocity of the initial Parker solution $(v_p(r))$ in the co-rotating frame by adding contributions from the planetary orbital motion, planetary rotation, and frame rotation. For simplicity we assume the planet is tidally locked so the frame rotation rate, orbital rotation rate and frame rotation rate are equal (we have only one value of $\Omega$). We also assume that the wind preserves its specific angular momentum as it spirals radially outward.  The wind velocity then reduces to
\begin{equation}
\boldsymbol{v}\left ( \boldsymbol{r} \right ) = \left \{
\begin{array}{cr}
        0 &  r \leq R_p \\
        v_p(r) \hat{\boldsymbol{r}} - \boldsymbol{\Omega} \times \hat{\boldsymbol{r}} \left ( r-R_{p}\right) & r > R_p \\
\end{array}
\right .
\label{eq:velocity_corrections}
\end{equation}
where $\boldsymbol{r}$ is the position vector relative to the planet. Thus with ISOROT and ANISOROT we can study the launching and flow dynamics in the vicinity of the planet with before moving on to global simulations.

Our fifth model ANISOROT* is a global simulation that includes both a stellar wind and the anisotropic planetary wind.  The stellar wind was modelled as a Parker wind emanating from the surface of the star at a temperature of $T_{\star} = 10^6 \mbox{K}$ corresponding to a $\lambda_{\star} = 22.09$.  The density at the stellar surface was set to be $\rho_{\star}=3.2112\times 10^{-13} \mbox{g cm}^{-3}$, a factor of $100$ times the density at the base of the planetary wind.  The velocity of both winds was then modified from the purely radial Parker wind solution by adding contributions from the stellar/planetary orbital motion, frame rotation, and stellar/planetary rotation as described in Equation \ref{eq:velocity_corrections}.  For convenience,  the stellar rotation was also set equal to the orbital rotation so that both the stellar and planetary boundaries were fixed in the co-rotating frame.  The star and planet's instantaneous orbital velocities and positions were used to calculate the various terms in Equation \ref{eq:velocity_corrections}.  This provides a fairly good approximation for the stellar wind since the dynamical propagation time of the stellar wind is much less than its orbital time scale and the orbital speed of the star is small compared to the speed of the stellar wind.  

The entire grid was initialized with this approximate solution for the stellar Parker wind, and then allowed to relax for an orbital time, while holding the solution fixed within the stellar launch region $r < 0.4 a$ and the outer boundaries.  Figure \ref{fig:parkerwind2} shows the density and radial velocity of the stellar wind perpendicular to the axis of rotation after allowing the wind to relax over one orbit period just prior to insertion of the planet.  

\begin{figure*}
\centering
\includegraphics[width=0.6\textwidth]{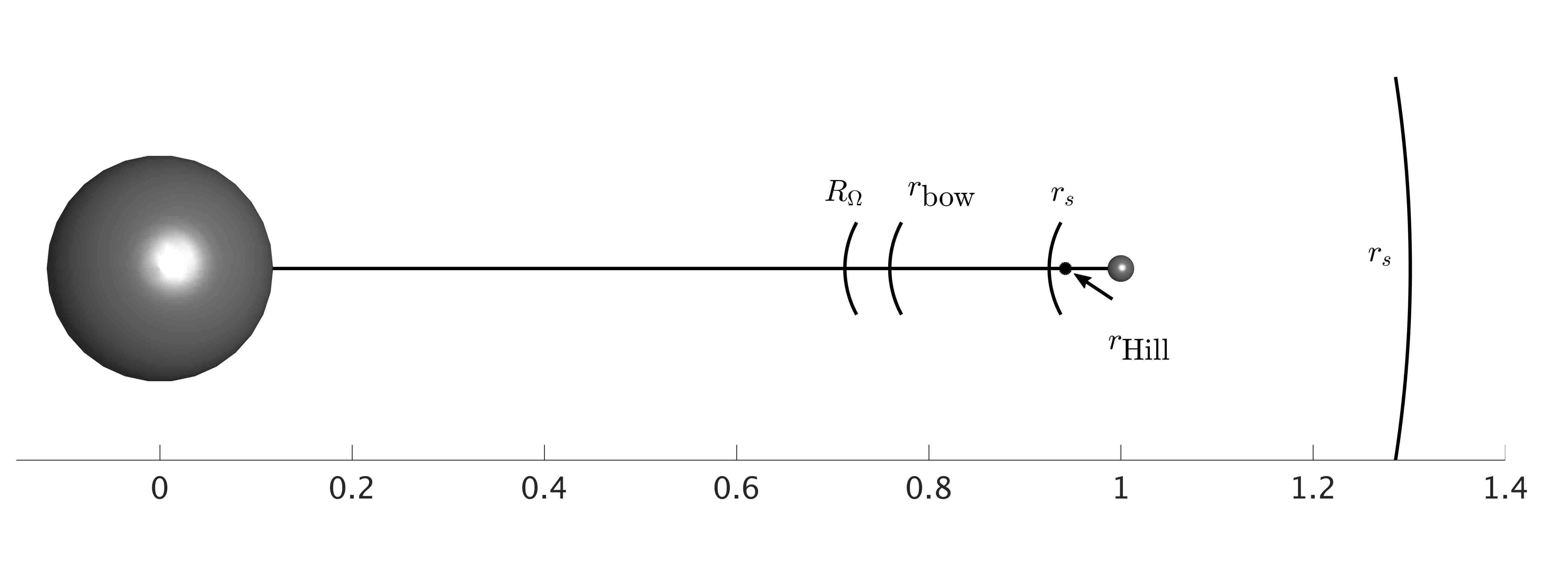}
\caption{Schematic showing relative sizes of star and planet as well as locations of sonic points $r_s$ for both stellar and planetary winds, the Hill radius $r_{\mbox{Hill}}$, Coriolis radius $R_{\Omega}$, and the bow shock radius $r_{\mbox{bow}}$}
\label{fig:1dschematic}
\end{figure*}

After allowing the stellar wind to relax for 1 orbit, we insert the planetary companion.  Similar to \cite{matsakos15} we estimate the distance at which the ram pressure of the planetary wind matches the ram pressure of the stellar wind.  We  then initialize the planetary Parker wind with the isotropic solution (taking into account planetary orbital motion, frame rotation, and planetary rotation) out to this radius $r_{\mbox{\small{bow}}} = 0.24 a$ where $a$ is the orbital separation.  This is well outside of the Hill radius of the planet $r_{\mbox\small{{Hill}}} = 0.058 a$ so we expect some material blown off by the planet to become gravitationally bound to the star.  This corresponds to a type III interaction as discussed in \cite{matsakos15}.  Figure \ref{fig:1dschematic} shows the relative sizes of the stellar and planet surfaces, the Hill radius, the bow shock radius, and the sonic points.  After inserting an isotropic Parker wind  we then continue to maintain the anisotropic planet boundary. This is then allowed to relax into a steady state over another 9 orbital timescales.

The ANISOROT* simulation was performed in a Cartesian grid co-rotating with the binary system about the z axis. The star and planet were located along the x axis at $\frac{-q}{q+1}a$ and $\frac{1}{q+1}a$ where $q=\frac{M_p}{M_{\star}}$. The simulation domain extended from $[-2,-2,-.5]\frac{a}{q+1}$ to $[3,2,.5]\frac{a}{q+1}$ with a base resolution of $320 \times 256 \times 64$.  An additional 4 levels of refinement were added giving an effective resolution of $5120 \times 4096 \times 1024$ which allowed for resolving the radius of the planet with 14 cells.  The other four non-global runs were run with the same setup and resolution (14 cells per planet radius), though without rotation, stellar gravity, or a stellar wind.  For those runs the simulation domain was reduced to a cube $0.25 \frac{a}{q+1}$ on a side centered on the planet.  This enables  following the planetary wind out to $9 R_p$ beyond the transition to a supersonic wind.

All our models are summarized in Table \ref{tab:runs}.


\begin{table}
    \centering
    \caption{Summary of Models}
    \label{tab:runs}
    
    \begin{tabular}{c|c|c|c}\hline
        Model & Planet wind & Co-rotating & Scale \\
        \hline
        ISO & Isotropic & No & Local \\
        ANISO & Anisotropic & No & Local \\
        ISOROT & Isotropic & Yes & Local \\
        ANISOROT & Anisotropic & Yes & Local \\
        ANISOROT* & Anisotropic & Yes & Global
    \end{tabular}

\end{table}

\section{Results}
We begin by focusing on the "local" simulations which focused on the near planet environment before moving to the global simulations which included both star and planet.

\subsection{Isotropic Planetary Wind}
Figure \ref{fig:parkerwind2} demonstrates the code's ability to recover the Parker type planetary wind by showing radial averages of the density and velocity for run ISO. 

\begin{figure*}
\includegraphics[width=0.85\textwidth]{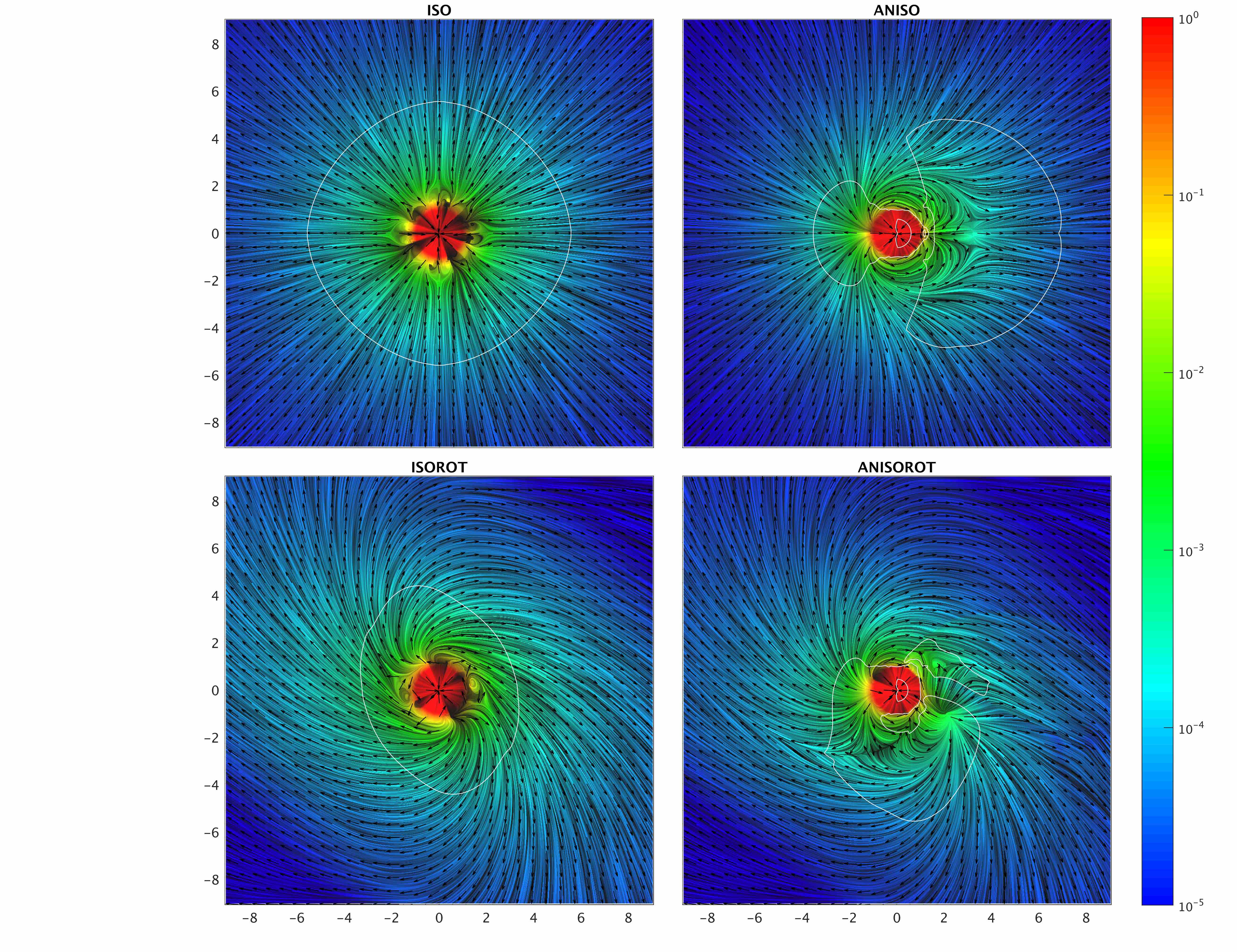}
\caption{Flow-texture plots of the four local runs showing density (hue) scaled to $\rho_p$ and local streamline orientation (texture).  The direction of the velocity field is shown by the quivers and the white contour shows the sonic surface.  The axis are in units of planet radii and for a pure Parker wind with $\lambda = \lambda_p$, this transition should occur at $\frac{\lambda_p}{2} R_p = 5.37 R_p$.}
\label{fig:launch}
\end{figure*}

The full multi-dimensional solution for the isotropic wind is shown in the top left panel of Figure \ref{fig:launch}. This flow-texture plot is constructed by performing line-integral convolution of a noisy black and white image  along the local velocity streamlines.  This effectively smears the image in the direction of the flow providing an accurate representation of the streamlines without having to select streamline "seeds". This black and white image is then colored using the density in the midplane.  The top left panel of Figure \ref{fig:launch} shows the generation of a spherical outflow passing through a spherical Mach surface.  Note that the location of the sonic radius is very close to $r_s = \frac{\lambda}{2} R_p$ as expected.The mass loss rate in the models is $9.65 \times 10^{10} \mbox{g s}^{-1}$ which is close to the theoretical mass loss rate of $10^{10}$ used to set $\rho_p$.

\subsection{Anisotropic Planetary Winds}
In the anisotropic model ANISO, the purely radial Parker wind solution is modified by pressure differences between the day side and night side of the planet.  These pressure differences drive strong lateral flows over the surface of the planet as seen in the upper right panel of Figure \ref{fig:launch}.

The lateral flows also change the location of the sonic surface.  Material that leaves the day side is now accelerated to supersonic velocities more quickly compared with the ISO case.  In addition, lateral pressure gradients drive the flow back towards the night side.  The convergence of the flow towards the night side produces a stagnation point that separates material re-captured by the planet from material that is accelerated outwards by a mix of pressure gradients and a reduced gravitational pull until it more closely resembles the Parker wind solution although with a slightly larger sonic radius.  Thus our solutions show the existence of a planetary backflow where material that was launched by the stellar UV flux in a wind on the dayside falls back onto the atmosphere on the night side.

\begin{figure}
\centering
\includegraphics[width=0.95\columnwidth]{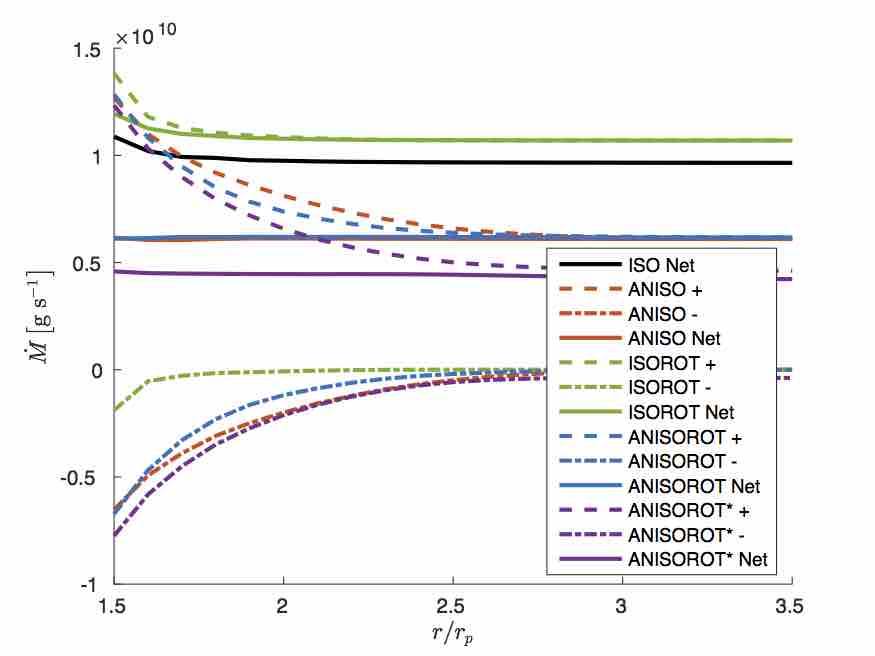}
\caption{Comparison of the total outward (+) and inward (-) mass flux through radial shells for all five models as well as the resulting net mass loss as a function of radial distance.  See Figure \ref{fig:massflux} for the angular distribution of mass flux at 2, 4, and 6 $r_p$.}

\label{fig:backflow}
\end{figure}

\begin{figure*}
\centering
\includegraphics[width=0.75\textwidth]{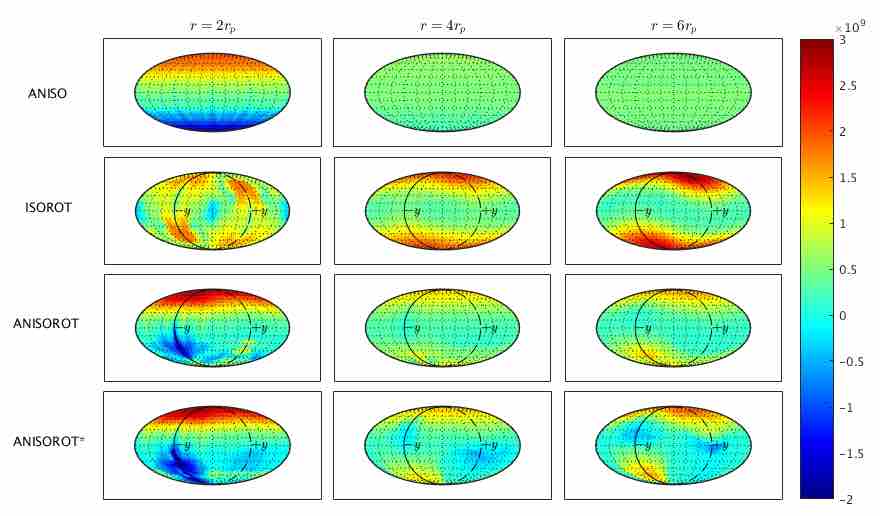}
\caption{Comparison of the planetary mass loss per solid angle in $\mbox{g s}^{-1} \mbox{sr}^{-1} $ measured at $2, 4, \mbox{ \& } 6$ planet radii for runs ANISO, ISOROT, ANISOROT, and ANISOROT*. The side of the planet facing the star is shown in the upper hemisphere and the leading edge of the planet in the models with rotation corresponds to $+y$ in this figure and the trailing edge to $-y$.}
\label{fig:massflux}
\end{figure*}

To quantify the magnitude of this backflow in the different solutions, in Figure \ref{fig:backflow} we present comparisons of the net mass loss as well as the total outward (+) and inward (-) masses flux through radial shells for all five models as a function of radial distance.   In addition, to see the geometry of flows more clearly, in Figure \ref{fig:massflux} we present the angular dependence of the mass flux sampled at shells of different radius for four of the simulations (ANISO, ISOROT, ANISOROT, ANISOROT*).

In the ANISO cases we see that close to the planet ($r=2r_p$; Figure \ref{fig:massflux}), a strong asymmetry develops as material leaves the day side and falls back on the night side. By $4 r_p$, however, the mass flux has become almost spherical again.  

Note also that the isotropic case has a mass flux per solid angle of $\frac{10^{10} \mbox{g s}^{-1}}{4 \pi} = 0.8 \times 10^9 \mbox{g s}^{-1} \mbox{std}^{-1}$.   This is less than  the value seen on the day side of the ANISO simulation $(1.87 \times 10^9  \mbox{g s}^{-1} \mbox{std}^{-1}$.   We attribute this difference to the fact that  material leaving the day side  can expand not only radially, but laterally as well.
While Figure \ref{fig:backflow} shows that the anisotropic case has a larger overall outflow close to the planet, the net outflow rate is reduced by $37\%$ to $6.1 \times 10^{9} \mbox{g s}^{-1}$. Note that by $R=3 r_p$ the flow is only outward.  We note that this ``backflow'' will have consequences in the full simulations (as described below) and may also have consequences for atmospheric dynamics.

\subsection{Isotropic Planetary Winds in a Binary System}
In the lower left panel of Figure \ref{fig:launch} we show the the density and streamline pattens for run ISOROT. The presence of the stellar gravity and the addition of the orbital motion has clearly changed the flow pattern near the planet and we see a reshaping of the flow into the beginning of two armed stream. 

The changes in flow pattern can be understood by recognizing that in a binary system, the planetary wind is subject to Coriolis and tidal forces in addition to gravitational and pressure forces.  The relative strength of these different forces will alter the flow structure of the Parker solution.  In appendix \ref{app:localforces}, we derive the equations of motion close to the planet in the limit $q << 1$ and $r_p << a$ where $\mathbf{r_p}$ is the relative position from the planet and $\mathbf{a}$ is the location of the planet relative to the star.  

\begin{figure}
\centering
\includegraphics[width=0.85\columnwidth]{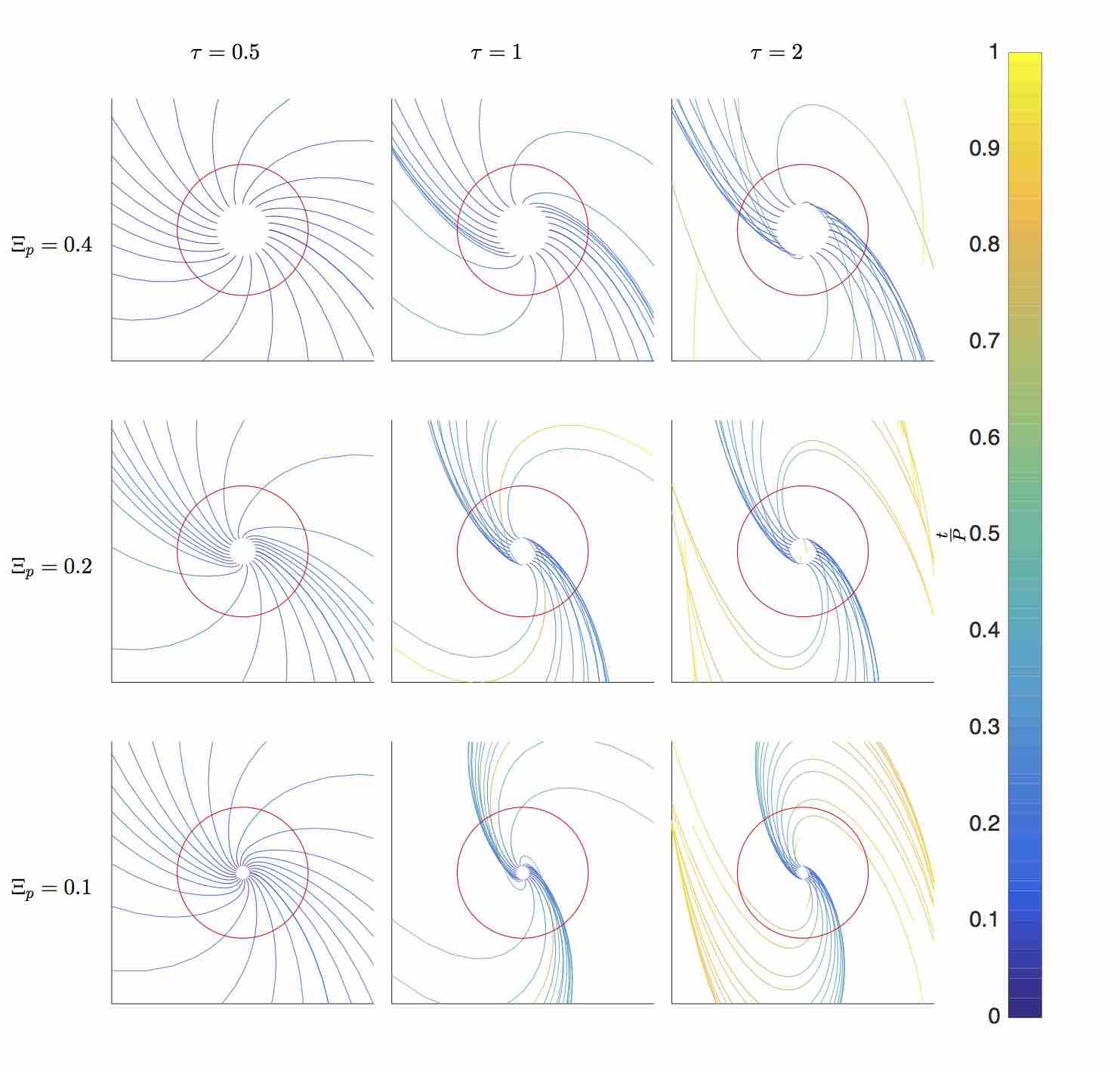}
\caption{Streamlines in the co-rotating frame for particles emitted from a planet surface for different values of $\Xi_p$ and $\tau$ in the limit of $r_s << a$.}
\label{fig:streamlineslocal}
\end{figure}

The combination of pressure gradients and gravitational attraction from the planet lead to an acceleration  given by the Parker solution (here $r_s$ is the sonic radius for the Planetary wind):
\begin{equation}
  a_P(r_p)=\frac{c_s^2}{2 r_s} \left . \frac{d\psi}{d\Xi} \right |_{\Xi=r_p/r_s}.
\end{equation}
The combined tidal and centripetal forces produce an accelerate of magnitude
\begin{equation}
 a_t(r_p)=3 \Omega^2 r_p,
\end{equation}
while the Coriolis force induced acceleration (assuming the Parker solution for the velocity) is of the order
\begin{equation}
  a_{\Omega}(r_p)=2\Omega v = 2 \Omega c_s \psi^{1/2},
\end{equation}
where the right hand expression is evaluated at $(\Xi=r_p/r_s)$. We can make these accelerations dimensionless by scaling these in units  of the Parker wind acceleration, ($c_s^2/r_s$), to get
\begin{equation}
\begin{aligned}
\alpha_P= & \frac{1}{2}\frac{d\psi}{d\Xi} \\
\alpha_\Omega = & 2 \tau \psi^{1/2} \\
\alpha_t = & 3 \tau^2 \Xi
\end{aligned}
\end{equation},
where $\tau = \frac{r_s \Omega}{c_s}$ is the ratio of the Parker time $\frac{r_s}{c_s}$ to the orbital time $\frac{1}{\Omega}$. To better understand the behavior of gas in the simulations we have carried out semi-analytic calculations of the streamlines under the influence of these accelerations. Figure \ref{fig:streamlineslocal} shows our results for different values of $\tau$ and $\Xi_p=\frac{2}{\lambda_p}=\frac{R_p}{r_s}$.

In general when $\tau << 1$ and the wind makes it to the transonic radius in less than an orbit, the subsonic regions of the Parker solution are largely unchanged.  The streamlines remain radial. For $\tau \ge 1$, tidal and Coriolis forces rotate and confine the  Parker solution streamlines into relatively narrow bands.  This trend can be seen across each column in Figure \ref{fig:streamlineslocal}.  In addition, the strength of the tidal, Coriolis, and pressure forces depend on $\Xi$ and the details of the solution will depend on the launch point for the wind within the atmosphere.  For smaller $\Xi_p$, meaning a longer relative distance between launch point and sonic radius, tidal forces dominate early on and the streamlines will be more compressed as they are rotated as shown in each column of Figure \ref{fig:streamlineslocal}.  

Note that ISOROT has $\tau = 1.2$ and $\Xi_p= 0.186$ and its velocity field shown in the bottom left panel of Figure \ref{fig:launch} resembles the center panel in Figure \ref{fig:streamlineslocal}.  Thus our analytic treatment of coriolis and tidal forces captures the behavior seen in the simulations.

Addressing the issue of backflow, we see that tidal and Coriolis forces drive wind inhomogeneities even in the isotropic launch case. Figure \ref{fig:backflow} shows a $10\%$ increase in the net outward mass flux in the ISOROT simulation as well as a small amount of inward mass flux near the planet surface for $r < 2 r_p$.   We attribute the increased net flux to tidal forces pulling material away from points near the sub-solar point and the opposite point on the night side (as seen in Figure \ref{fig:massflux}).  Figure \ref{fig:massflux} also shows a small degree of inward flux at $r = 2 r_p$ concentrated near the points above and below the orbital plane where gravity from the primary and the Coriolis force cause material to fall back onto the planet boundary.  At larger radii we see the formation of two streams that are then bent by the Coriolis force.  The stream on the day side is shifted "up-orbit" (towards +y) and the stream leaving the night side is shifted "down orbit" (towards -y).  

\subsection{Anisotropic Planetary Winds in a Binary System}
Finally we consider how day/night asymmetry in the wind launching   affects the flow in the regions close to a planet on a hot orbit. 

In the lower right panel of Figure \ref{fig:launch} we see the resulting flow structure.  The flow coming from the day side (-x) is similar to that seen in ISOROT, however, the stagnation point behind the planet (+x) where the lateral flows converge is shifted down-orbit (-y) by the Coriolis force compared to the same point in run ANISO where orbital motion was not considered.  This downorbit shift is one aspect of a more complex flow pattern near the planet occuring in the ANISOROT case relative to the ISOROT simulation.  

Note that we see velocity vectors diverging in the orbital plane. This occurs due to convergence from above and below.  The resulting backflow is then turned slightly upward and runs into material travelling around the leading edge (+y) of the planet. Material that is pushed outward from this convergence point is also bent by the Coriolis force until it impacts material coming around the trailing edge (-y) forming a secondary stagnation point and a second backflowing stream that merges with the first backflow.

In Figure \ref{fig:backflow} we see that rotation does not significantly increase the net mass flux, unlike what was seen in the isotropic case.  We attribute this change to tidal forces competing against the pressure gradient driving the wind.  Together this leads to higher densities and pressures above the day side compared to the non-rotating case (ANISO) and a decreased overall outward flux.  

The panel corresponding to $r=2r_p$ for run ANISOROT in Figure \ref{fig:massflux} shows the backflow as in the ANISO case, but now shifted down-orbit by the Coriolis force.  At larger radii, the flow resembles the two stream structure of ISOROT, but with only $57\%$ of the net flux due to the reduced temperature on the planet's night side.  The boundaries in run ANISOROT prevented material from re-entering the grid.  In our global run ANISOROT*, discussed in the next section, we were able to follow the large scale evolution of the flow.  In that simulation some material sent up-orbit eventually turns around and falls back towards the night side of the planet.  This increased pressure reduces the overall outward mass flux by $30\%$ to $4.25 \times 10^{10} \mbox{g s}^{-1}$ compared to ANISOROT as can be been seen by comparing the runs in Figures \ref{fig:backflow} and \ref{fig:massflux}


\subsection{Large scale behavior of Planetary winds}

\begin{figure}
\centering
\includegraphics[width=\columnwidth]{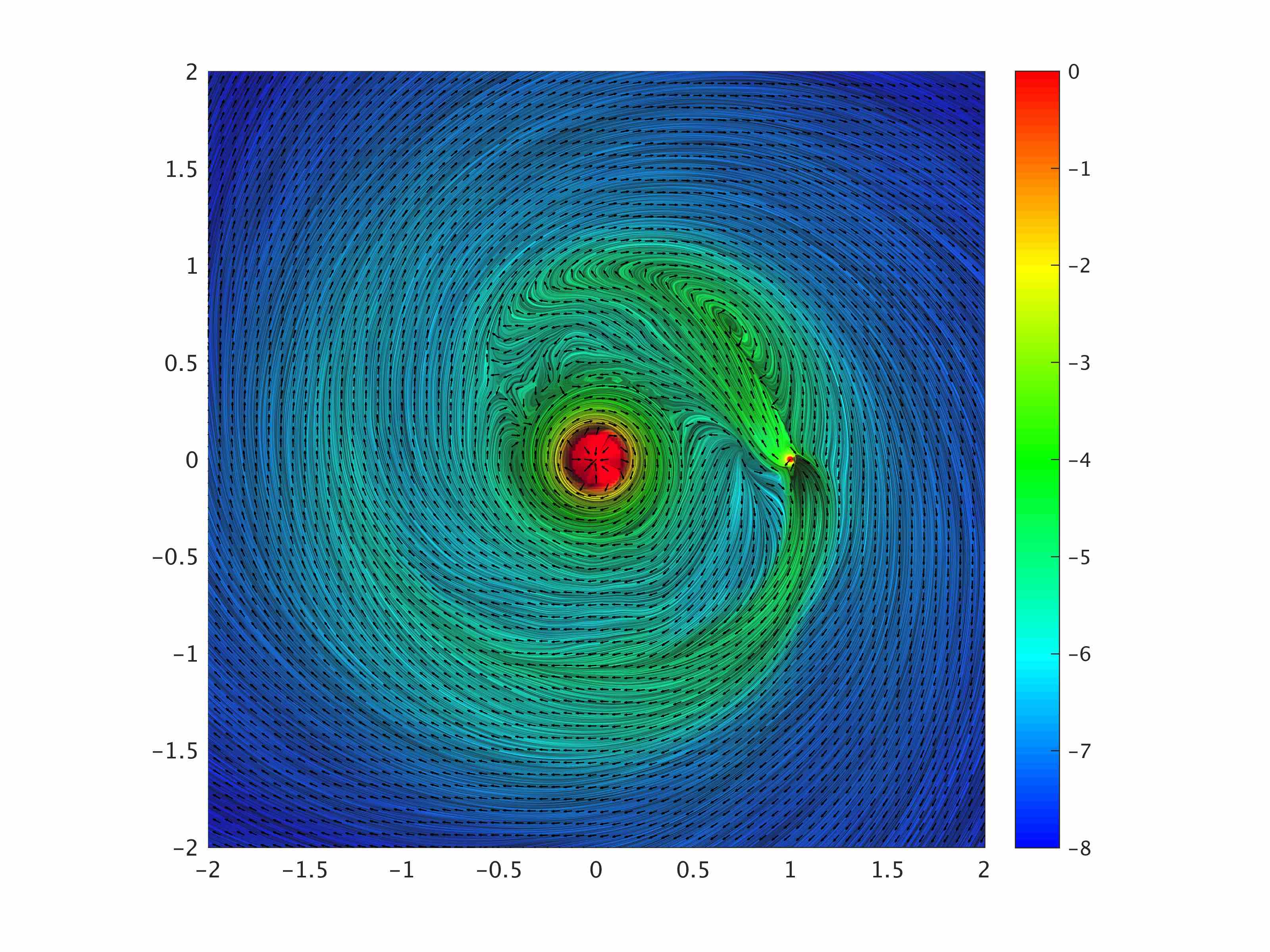}
\includegraphics[width=\columnwidth]{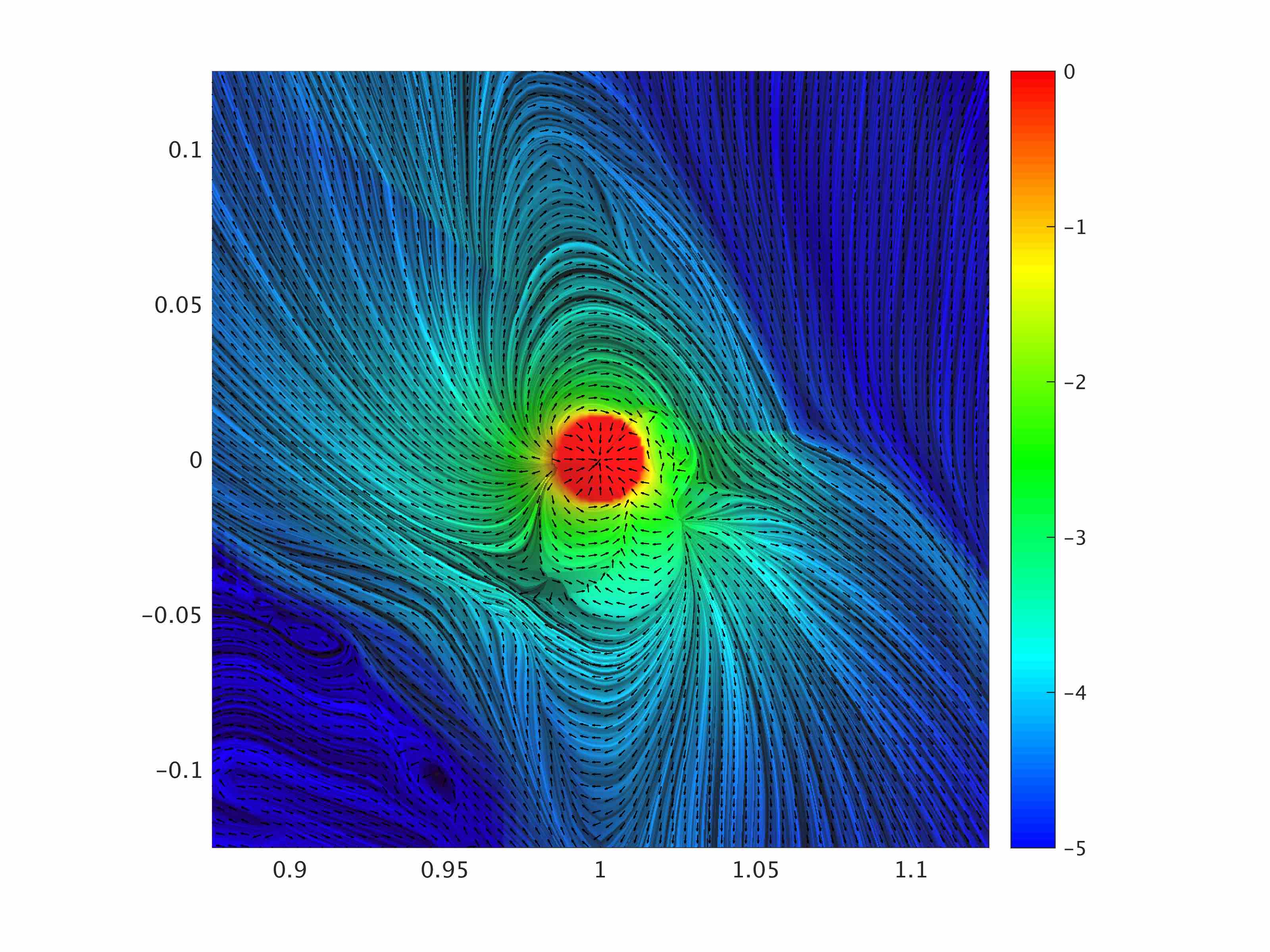}
\caption{Flow-texture plot of time averaged velocity field in orbital plane.  The time average was generated using 50 snapshots over the last 5 orbits.  The hue correponds to the density in units of $\rho_p$ and the direction of the velocity field is shown both by the texture of the image as well as the velocity vectors}
\label{fig:flowfield}
\end{figure}

In top panel of Figure \ref{fig:flowfield} we show the large scale flow pattern for the global simulation ANISOROT*.  The most prominent feature in the run is the presence of the of the up-orbit and down-orbit streams.  The initial redirection of the planetary wind into the two streams was apparent in our local simulations discussed above and in our semi-analytic calculation of near-planet gas parcel trajectories.  Such
streams were also seen in the global simulations \citep{matsakos15}.  Note that on larger scales we see the density of the streams is considerably higher than the $r^{-2}$ fall off expected from a spherically symmetric outflow.  We find material with $\rho \sim 10^{-18} \mbox{g cm}^{-3}$ extending almost $90^\circ$ up orbit from the planet and almost $180^\circ$ down orbit. 

The bottom panel of Figure \ref{fig:flowfield} shows the effect of the temperature asymmetry on the global flow near the planet. The majority of the streamlines emerge from the day side of the planet where heating is expected to produce exobase temperatures of $T_p \sim 10^4$ K.  The streamlines from this hemisphere are redirected into the leading and trailing arms of the flow.  The planet's night-side, however, contributes very little to global flow.  As discussed in previous sections a backflow is established due to the asymmetry in the day/night side launching. In this global simulation we also see the effect of the stellar wind in the compression of the two stream pattern as the stellar wind flows over and around the planetary material. 

In what follows we expand the analysis carried out in the previous sections to explicate the origin and nature of the global two stream pattern in terms of the forces acting on gas parcels and their impact on the orbital trajectories.

On small scales, we considered the limit of $r_s << a$, and looked at how the ratio of the Parker time $\frac{c_s}{r_s}$ to the orbital time $\Omega^{-1}$ as well as the location of the planet surface relative to the sonic radius $\Xi_p = \frac{2}{\lambda}$ determined the evolution of the flow out to the planetary sonic point (Figure \ref{fig:streamlineslocal}).  In particular changing these ratios changed how asymmetric and how tilted streamlines became as they passed the sonic surface. Beyond the sonic surface, however, gravitational and pressure forces from the planet decrease.  Material that escapes the planet's gravity goes into orbit about the star.  The position and direction of gas parcels as they leave the planet's sonic surface determine their orbit. 

To better understand the large scale flow pattern in run ANISOROT* (seen in Figure \ref{fig:flowfield}), consider the ballistic trajectories of particles travelling radially outward from the sonic surface at a velocity $c_s$ in the limit that $q << 1$.  The position $(\mathbf{r'})$ and velocity $(\mathbf{v'})$ of each particle in the inertial frame is given by 
\begin{equation}
\begin{aligned}
    \mathbf{r}'= &\mathbf{a} + r_s \mathbf{\hat{n}} \\
    \mathbf{v}'= & \mathbf{\Omega} \times \mathbf{r}' + c_s \mathbf{\hat{n}}.
\end{aligned}
\end{equation} 
where $a$ is the displacement vector from the star to the planet, or equivalently the location of the planet from the center of mass in the limit $q << 1$, $r_s$ is the sonic radius of the planetary wind, and $\mathbf{\hat{n}}$ is the unit vector normal to the sonic surface.  In the limit of  $c_s << v_p$ and $r_s << a$, we can express the semi-major axis of the resulting orbit of each particle as (see Appendix \ref{app:orbitaltrajectories})
\begin{equation}
    \frac{a'}{a}=1 +  2 \xi_T \sin \theta \sin \left ( \phi - \phi_c \right)
\end{equation}
where $\phi$ is the angle measured from $\mathbf{\hat{a}}$ around $\mathbf{\hat{\Omega}}$. Note that trajectories leaving the planet from its leading hemisphere (the one facing up orbit) have $0 < \phi < \pi$ . The angle $\theta$ is the angle measured down from the axis of rotation $\mathbf{\hat{\Omega}}$ ($\theta = \pi/2$ in the orbital plane). 


In addition, 
\begin{equation}
\begin{aligned}    
\xi_T = & \sqrt{\frac{c_s^2}{a^2 \Omega^2}+4 \frac{r_s^2}{a^2}} \\
\phi_c = & \arctan\left (-2 \frac{r_s\Omega}{c_s} \right ) = \arctan \left ( -2 \tau \right ) 
\end{aligned}
\end{equation}
where $-\pi/2 < \phi_c < 0$. As we will see, the angle $\phi_c$ splits the planet into hemispheres driving up-orbit and down-orbit arms respectively

When $\tau << 1$, so that the wind acceleration time is much less than an orbital period, we have $\phi_c = 0$.  In this case particles leaving the planet's leading hemisphere $(0 < \phi < \pi)$ have $a'> a$.  These gas parcels have longer orbital periods than the planet.  Thus they must have their trajectories turned around to form the down-orbit stream.  Particles leaving the trailing hemisphere ($\pi < \phi < 2 \pi$) have shorter orbits than the planet and form the up-orbit stream.  At the great circle that passes through $\phi = \phi_c$, material leaves on an elliptic orbit with the same orbital period as the planet.  These streamlines bend back around without moving up or down stream .  

\begin{figure*}
\centering
\includegraphics[width=0.85\textwidth]{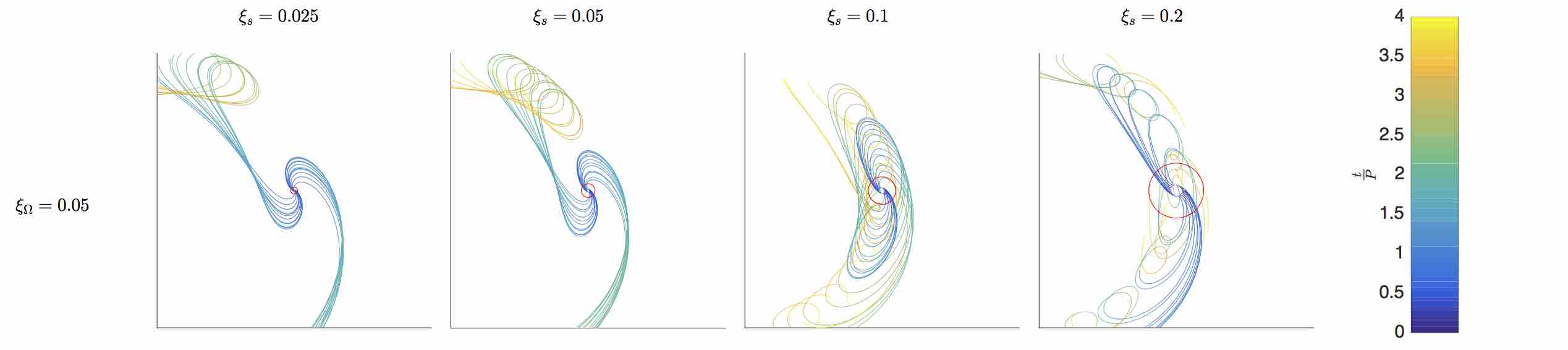}
\caption{Streamlines in the co-rotating frame for particles emitted from a planet surface for different values of $\xi_s$ with $\xi_\Omega=.05$ and $\Xi_p=.2$}
\label{fig:streamlinesglobal}
\end{figure*}

Thus as we increase $\tau$, material leaving near $\phi_c$ (or $\phi_c + \pi$) shifts from forming the up-orbit stream to forming the down-orbit stream (or vice-versa).  To explore this effect in more detail, we once again compute trajectories semi-analytically. Figure \ref{fig:streamlinesglobal} shows results for range of values of $\xi_s = \frac{r_s}{a}$.  These trajectories are computed using a fixed $\Xi_p=0.2$ which corresponds to $\lambda=10$ and $\xi_{\Omega}=\frac{c_s}{a \Omega} = 0.05$.

Since we fix $\tau$ in these plots, $\phi_c =-76^\circ$ is constant.  Thus in all the plots we see material leaving the planet being focused into up-steam or down-streams depending on their origin angle relative to $\phi = \phi_c$ and $\phi_c+\pi$.  Note that for ANISOROT*, $\xi_\Omega=\frac{c_s}{a \Omega}=0.062$, $\xi_s =\frac{c_s}{a \Omega} = 0.0740922$ and $\phi_c = -67.3^{\circ}$. 
Although the flow is modified by the anisotropic planetary wind, pressure gradients, and ram pressure from the stellar wind, we see the simulation does resemble its closest semi-analytic model.

It is perhaps more natural to consider the trajectories of the planetary wind in the co-rotating frame where streamlines are bent by the Coriolis force $-2\mathbf{\Omega} \times \mathbf{v}$. This bending can be seen in Figure \ref{fig:streamlinesglobal}.  For particles moving at a constant velocity $c_s$, the diameter  of curvature $R_\Omega$ scaled to the orbital separation $a$ is given by 
\begin{equation}
 \xi_{\Omega} = \frac{R_{\Omega}}{a} = \frac{c_s}{a \Omega} = \sqrt{\frac{GM_p}{\lambda R_p a^2}\frac{a^3}{G\left(M_p+M_s \right) }} = \sqrt{\frac{aq}{\lambda R_p \left(q+1 \right)}}
\label{rc}
\end{equation}
The final expression is described in terms of the mass ratio $q = M_p/M_{\star}$, the planet radius $R_p$, the orbital radius $a$, and the hydrodynamic escape parameter $\lambda$. 

From this expression we can see that for a fixed planet mass and radius, the distance that must be traversed before the Coriolis force can significantly influence the orbits increases with increasing orbital separation.  For $R_{\Omega}/a \ge 1$ we would expect the streamlines to return to a spherical configuration as is expected for an isolated Parker-type wind.  Thus it is only for planets on hot orbits where $R_{\Omega}/a < 1$ that significant streamline distortion can be expected.  

The dependence of Eq. (\ref{rc}) on $q^{1/2}$ for low $q$ shows that even for small orbital separations, strong Coriolis bending will only occur for small companions (i.e. planets). Thus we would not expect much deflection of winds to occur in binary stars once $q$ becomes of order 1.

As noted above strong distortion of the streamlines driven by the Coriolis and tidal forces leads to much higher densities at large distances from the planet than would be expected for purely spherical wind ($\rho \sim r^{-2}$).  In the simulations pressure forces (and resulting shocks) lead to bounded streams in which the planetary wind is redirected into parallel flowing up-orbit and down-orbit arms by tidal forces.  In general the material that forms the up-orbit and down-orbit streams will respectively converge into circular orbits while conserving angular momentum.  Using this we can estimate the angular drift per orbital period $\Theta_D$ from the planet for the up-orbit and down-orbit streams.

This gives (see appendix \ref{app:orbitaltrajectories})
\begin{equation}
    \Theta_D = 2 \pi \left ( \Omega'-\Omega \right ) = \mp 3 \left ( -4 \xi_s \sin \phi_c + 2 \xi_{\Omega} \cos \phi_c \right )
\end{equation}

In the limit that $\xi_s << \xi_{\Omega}$, $\phi_c << 1$ we have $\Theta_D = 6 \xi_{\Omega}$.  For run ANISOROT* this gives $\theta_D = \pm 38.7^\circ$ consistent with the large eddy seen in the top panel of Figure \ref{fig:flowfield}.

\begin{figure*}
\centering
\includegraphics[width=0.7\textwidth]{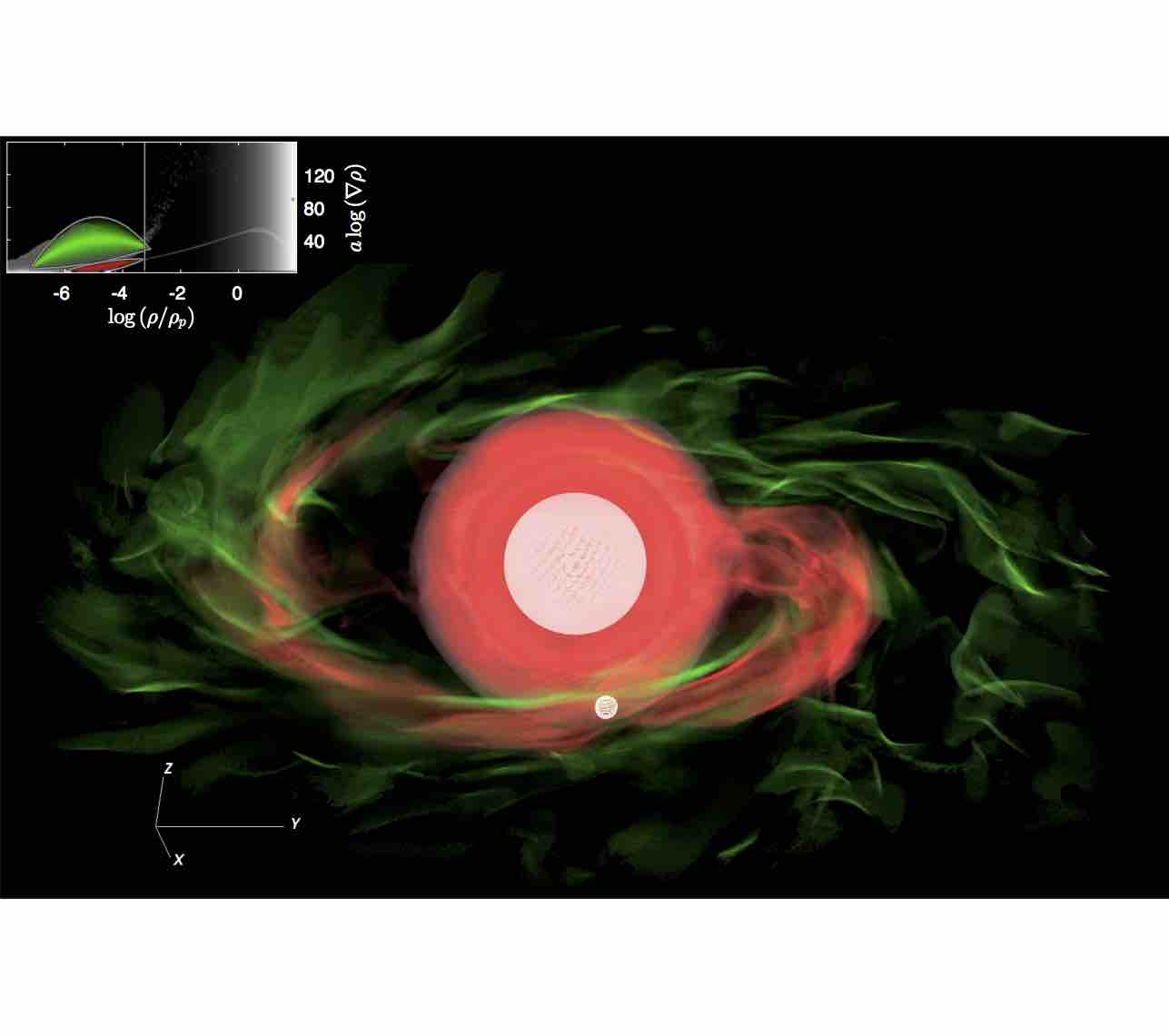}
\caption{3D volume rendering of the ROT model after 10 orbits using SLIVR (a hardware accelerated volume rendering library developed at the SCI Institute at the University of Utah) with the 2D transfer function shown on the top-left corner. The star and planet are both shown in white.  The transfer function hides most of the unperturbed stellar wind seen as the thin line in the transfer function histogram. The green shows material with a larger gradient than the stellar wind corresponding to the interfaces between hot and cold material as well as any shocks, while the red shows a portion of the stellar atmosphere as well as smoother parts of the planetary flow.  The camera is located $27^{\circ}$ above the orbital plane.}

\label{fig:3Dvol}
\end{figure*}

To gain a more complete view of the flow in Figure \ref{fig:3Dvol} we present a 3-D volume rendering of the simulation.  This image views the flow from angle of $27^{\circ}$ above the orbital plane.  We use a 2-D opacity function in which one dimension is sensitive to density and the other is sensitive to gradients in density. The opacity function is shown in the upper left. This allows the image to capture the location of complex shock structures. 

Once again, the forward (up-orbit) and backward (down-orbit) streams are the most apparent feature of the simulation.  A number of important, smaller scale flow features become noticeable however.  Note the that flow near the planet is highly asymmetric.  In addition we see that forward (up-orbit) arm shows significant fragmentation at it's terminus. We also see an apparent inflow onto the star from the inner edge of the forward stream.  Thus while the overall 2-arm behavior of the flow is relatively stable over many orbits we do see smaller scale structure in flows whose time-dependence we address in the next section.

\subsection{Time Dependence}
\begin{figure*}
\centering
\includegraphics[width=.8\textwidth]{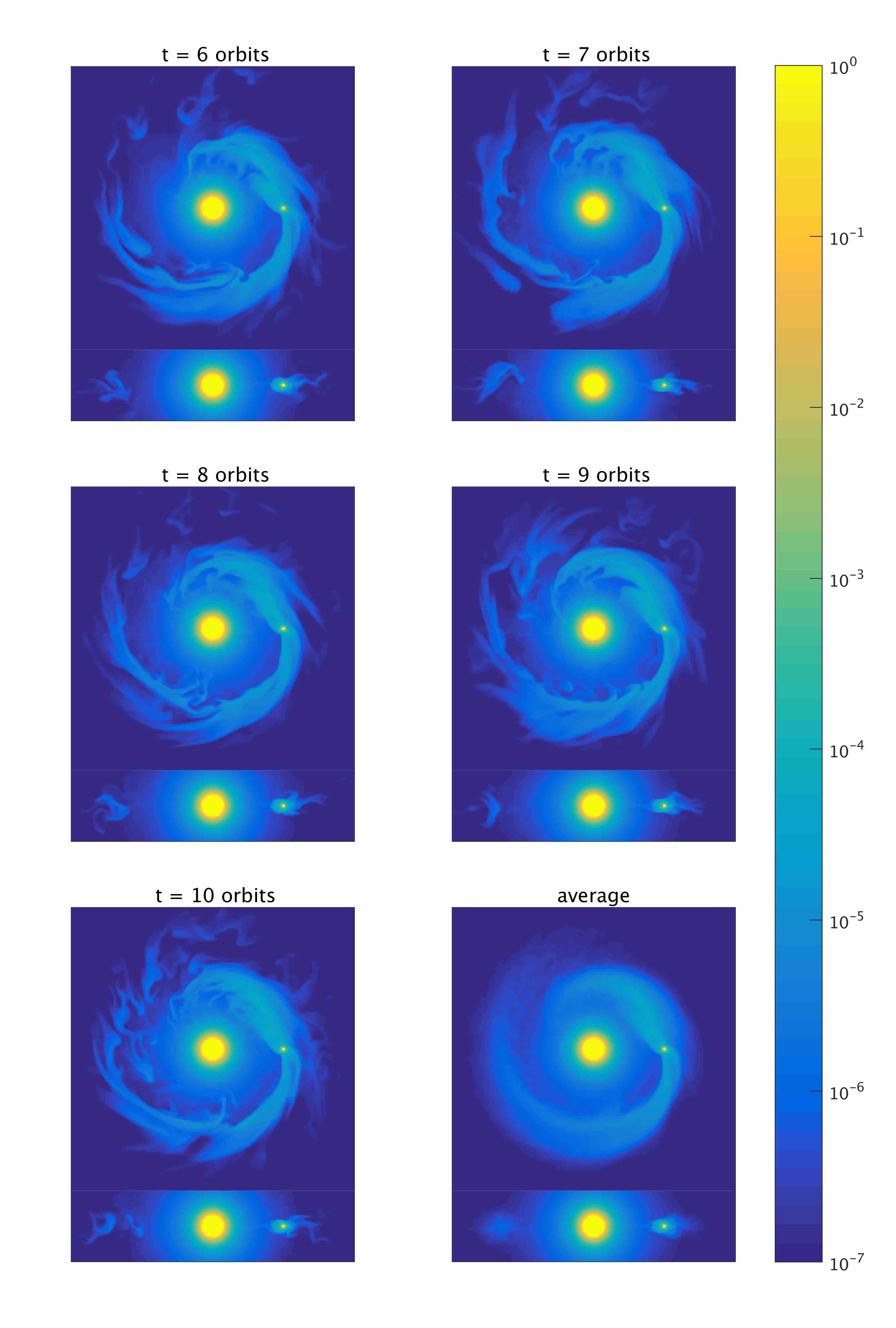}
\caption{Slices of density in both the orbital plane at $z=0$ (where the orbit is counter-clockwise) and perpendicular to it (at $y=0$) of the co-rotating model (ROT) scaled to $\rho_p$.  The planetary wind was turned on after 1 orbital period and the simulation was allowed to relax for another 5 orbital periods.  The snapshots above were then taken at $t=6, 7, 8, 9, \mbox{ \& } 10$ orbital periods $P$ and the average was produced by taking 50 snapshots over the same time frame.}

\label{fig:series}
\end{figure*}

In Figure \ref{fig:series} we present snapshots of density of the system in, and perpendicular to, the equatorial plane at 5 different times.  We also present the image made by averaging over the 5 frames.  Taken together, these plots allow us to focus on details of the global interaction between stellar outflow and planetary wind.  In particular we see significant differences between the leading and trailing arms of the planetary wind flow.

First note from Figure \ref{fig:series}, that the density in the flow falls of sharply within a $R \sim 2 - 3 R_p$ perpendicular to the orbital plane.  The reason for this is a combination of the Coriolis force (as discussed in the last section) as well as the hydrodynamic interactions of the planetary wind with the stellar outflow.  In contrast to the familiar bow shock which forms for planets at larger distances, here the stellar outflow sweeps around the cylindrical column of the leading and trailing arms of the planetary wind (see Figure \ref{fig:3Dvol}).  

In the orbital plane at larger distances (and hence angles) from the planet we see the presence of instabilities along both leading and trailing arms of the planetary wind flow.  A complex, time-dependent flow emerges from the interaction between the planetary wind and stellar wind.  This can be seen along the inner edge of both the up-orbit and down-orbit streams. In addition note the termination and mixing in the regions near the head of the leading stream.  Consideration of animations of the simulations shows the structures seen there form as a material is stripped off the leading stream through its interactions with the stellar outflow and fall back towards the star.  

\section{Observational Consequences}

\subsection{Synthetic Observations}
In order to generate synthetic observations, we calculate the optical depth per frequency as 
\begin{equation}
    \tau \left( \nu \right ) = \displaystyle \sigma_{\nu_0} \int { ds \int { d\nu' n \left( \nu' \right ) \phi \left ( \nu - \nu' \right ) } }
\end{equation}
where $\sigma_{\nu_0} = 1.102 \times 10^{-2} \mbox{cm}^2 \mbox{s}^{-1}$ \citep{bourrier13} and the line integration is along rays emanating from the surface of the star towards an observer.  Since each voxel has a particular velocity, the distribution of density per frequency for cell $j$ is just 
\begin{equation}
 n\left (\nu' \right ) = n_j \delta \left ( \nu'-\nu_j \right ) \mbox{ where } \nu_j = \nu_0 \left ( 1 - \frac{\mathbf{v}_j \cdot \hat{n}}{c} \right ),
\end{equation}
where $n_j$ is number density of cell $j$.
This gives us
\begin{equation}
    \tau \left( \nu \right ) = \displaystyle \sigma_{\nu_0} \int { ds \, n_j \phi \left ( \nu - \nu_j \right ) }
\end{equation}
Then if we integrate over frequency bin $i$, we get
\begin{equation}
    \tau_i \Delta \nu = \displaystyle \int_{\nu_i-\frac{\Delta \nu}{2}}^{\nu_i+\frac{\Delta \nu}{2}} \tau \left ( \nu \right ) d \nu =  \sigma_{\nu_0} \int { ds \, n_j \int _{\nu_i-\frac{\Delta \nu}{2}}^{\nu_i+\frac{\Delta \nu}{2}} \phi \left ( \nu - \nu_j \right ) }
\end{equation}

We assume that the line profile is from thermal broadening and estimate it as 

\begin{equation}
    \phi \left ( \nu - \nu_j \right) = \frac{1}{\sqrt{\pi} \Delta \nu_D} \exp^{-\left ( \nu-\nu_j \right )^2/\left ( \Delta \nu_D \right ) ^2 }
\end{equation}
where $\Delta \nu_D = \frac{\nu_0}{c} \sqrt{\frac{2 k_B T}{m}}$

The integral then becomes
\begin{equation}
    \tau_i =  \frac{\sigma_{\nu_0}}{2\Delta \nu } \int { ds \, n_j \left [ \erf{\left ( \frac{ \nu_j - \nu_i + \frac{1}{2} \Delta \nu}{\Delta \nu_D} \right )} - \erf { \left ( \frac{\nu_j-\nu_i-\frac{1}{2} \Delta \nu}{\Delta \nu_D} \right ) } \right ] }
\end{equation}

This can also be defined by an effective absorption coefficient per cell $j$ 
per frequency bin $i$ of
\begin{equation}
    \alpha_{j,i}  =  \frac{n_j\sigma_{\nu_0}}{2 \Delta \nu} \left [ \erf{\left ( \frac{ \nu_j - \nu_i + \frac{1}{2} \Delta \nu}{\Delta \nu_D} \right )} - \erf { \left ( \frac{\nu_j-\nu_i-\frac{1}{2} \Delta \nu}{\Delta \nu_D} \right ) } \right ]
\end{equation}
which can be integrated along rays emanating from the surface of the star using standard ray tracing techniques
\begin{equation}
  \tau_i = \int{\alpha_{j,i}\, ds}.
\end{equation}

\begin{figure}
\centering
\includegraphics[width=\columnwidth]{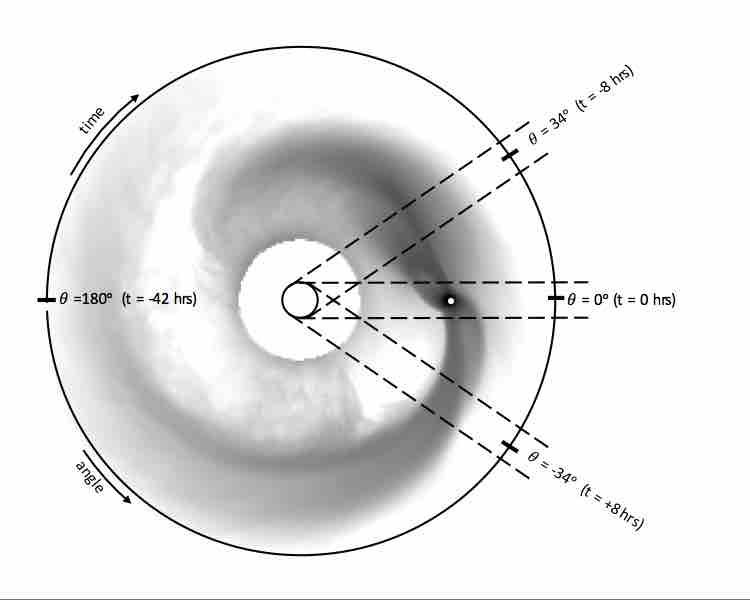}
\caption{Schematic showing top down view of orbital plane and relationship between instantaneous viewing angle and corresponding theoretical observational time.  The white region in the middle shows the stellar wind boundary and the stellar surface is shown by the black circle within. Synthetic absorption profiles calculated for different camera angles are shown in Figure \ref{fig:transit}.}
\label{fig:schematic}
\end{figure}

The simulation data was saved every 1/10th of an orbit. For each frame, sets of rays emanating from the surface of the star were integrated through the simulation domain at different angles to create synthetic images with a resolution of $512^2$ at each frequency and angle of observation.  The frequency resolution corresponds to $20 \mbox{ km/s}$ and the angle resolution corresponds to $10 \mbox{ min}$ of orbital time. These images were then summed to calculate a normalized intensity as a function of frequency, angle, and simulation time.  To the extent that the fluid flow is steady, different camera angles for a fixed simulation time can be translated to different observational times.  The relationship between angle and corresponding observational time are shown in Figure \ref{fig:schematic}.  Finally for each frequency and camera angle, we calculated the mean and standard deviation of the normalized intensity over the last 5 orbits.  

\begin{figure}
\centering
\includegraphics[angle=0,width=\columnwidth]{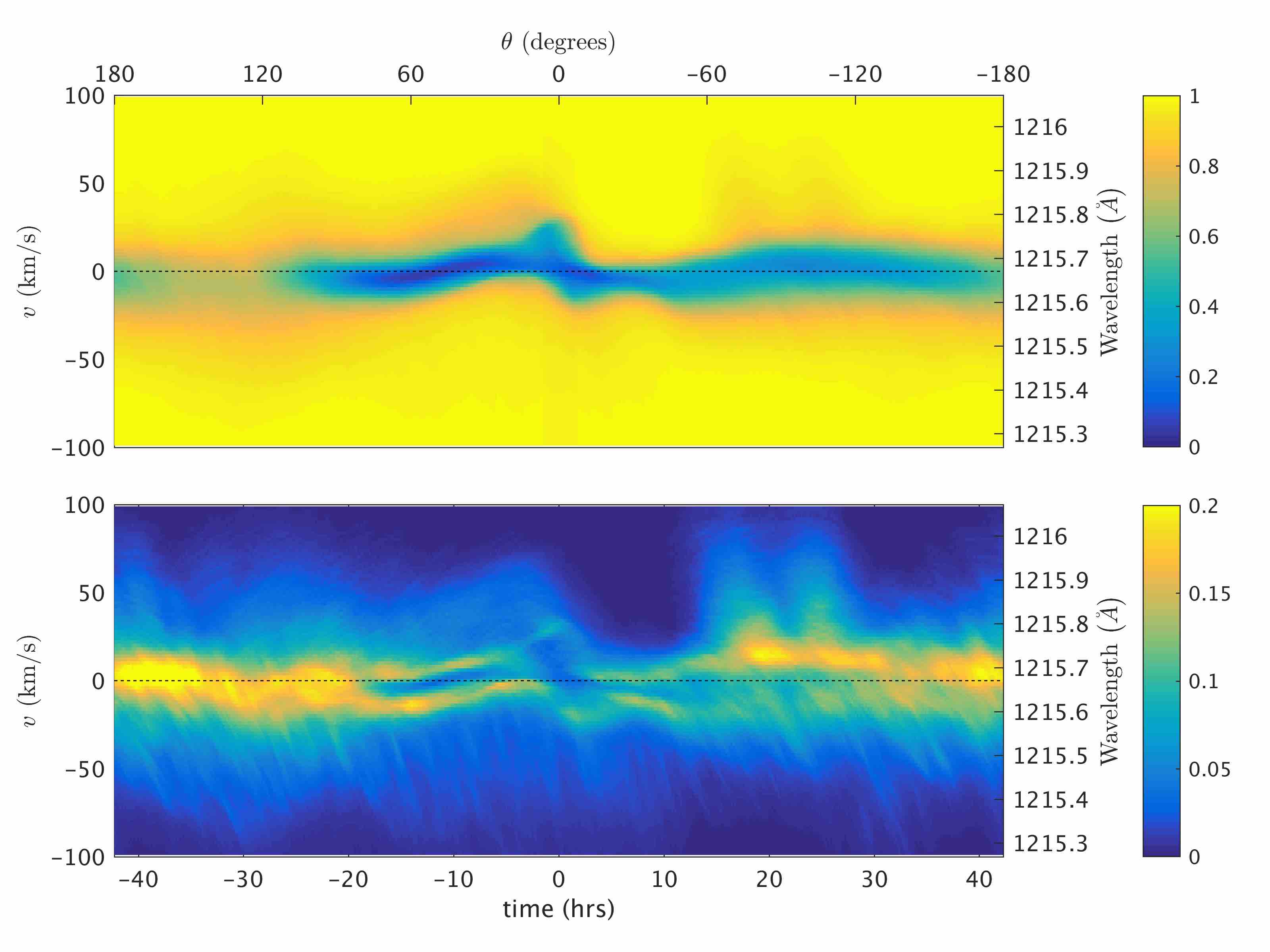}
\caption{The top panel shows average attenuation over the final 5 orbits as a function of line of sight velocity  (wavelength) and camera angle (transit time).  Positive velocities are towards the star.  The bottom panel shows the standard deviation of the attenuation over the final 5 orbits.  See Figure \ref{fig:schematic} for a schematic showing the relationship between time in the lab frame, and angle in the corotating frame.}
\label{fig:transit}
\end{figure}

\subsection{Variability}
The mean of the normalized intensity as a function of camera angle (or observational time) and wavelength (or line of sight velocity) is shown in the top panel of Figure \ref{fig:transit}.  The planet is immediately in front of the camera located at 0 degrees (corresponding to a transit time of 0 hrs) and absorbs light at all wavelengths during the transit $(\pm 1.5 \mbox{hrs})$ corresponding to $\pm 6^{\circ}$.  Were it not for the planets small radius, it would be visible as a vertical bar in the center of the top panel.  As we go a few degrees up-orbit (to the left) corresponding to earlier observational time, we see a trend towards larger positive line of sight velocities.  This is due to material being launched from the day side with positive line of sight velocities and then being bent up-orbit due to the Coriolis force.  The opposite trend is seen as we go a few degrees down-orbit (to the right), where we see larger negative line of site velocities caused by material leaving the planet, heading away from the star, and then being bent down-orbit by the Coriolis force.  

At larger angles we see a reversal in the line of sight velocities with material now heading away from the star at $\theta > +60 ^{\circ}$ and material now heading towards the star at $ \theta < -60 ^{\circ}$.  This we attribute to the streams being on slightly elliptic trajectories which would oscillate on the order of the drift angle per period $\Theta_D = 42^{\circ}$. 
In addition, ram pressure from the stellar wind pushes material radially outwards from the star towards negative line of sight velocities.  For the up-orbit stream, this ram pressure works with the Coriolis force to drive the large eddy seen in Figure \ref{fig:flowfield} centered at approximately 45 degrees, while for the down-orbit stream, the ram pressure works against the Coriolis force and resists the flow from turning inward and forming a second eddy down-orbit.  

As we continue further up-orbit, past the large eddy, we see a gap in the absorption between $120^{\circ}$ and $170 ^{\circ}$ where material sent up-orbit has either been accreted or been blown back and outward by ram pressure while material sent down-orbit has been diffused and ionized. This same gap is visible in Figure \ref{fig:schematic} where we show the neutral hydrogen column density normal to the orbital plane.

The bottom panel of Figure \ref{fig:transit} shows the standard deviation of the normalized intensity as a function of frequency and time (camera angle).  In general, we see increased variability as we head away from the planet in both the down and up-orbit directions.  There is also a trend towards increased variability at more positive line of sight velocities in the down-orbit stream.  This we attribute to the interaction of the stellar wind with the down-orbit stream producing Kelvin-Helmholtz and Rayleigh Taylor instabilities that cause some material to loose angular momentum and fall inward onto the stellar boundary.  This effect is visible for both the down-orbit and up-orbit streams in Figure \ref{fig:series}.  We also see large deviations in the intensity of material with larger negative line of sight velocities.  This we attribute to material that has been accelerated outwards by ram pressure over $\approx 1-2 $ orbits.  This same material is shown in green in Figure \ref{fig:3Dvol}.

\begin{figure}
\centering
\includegraphics[width=\columnwidth]{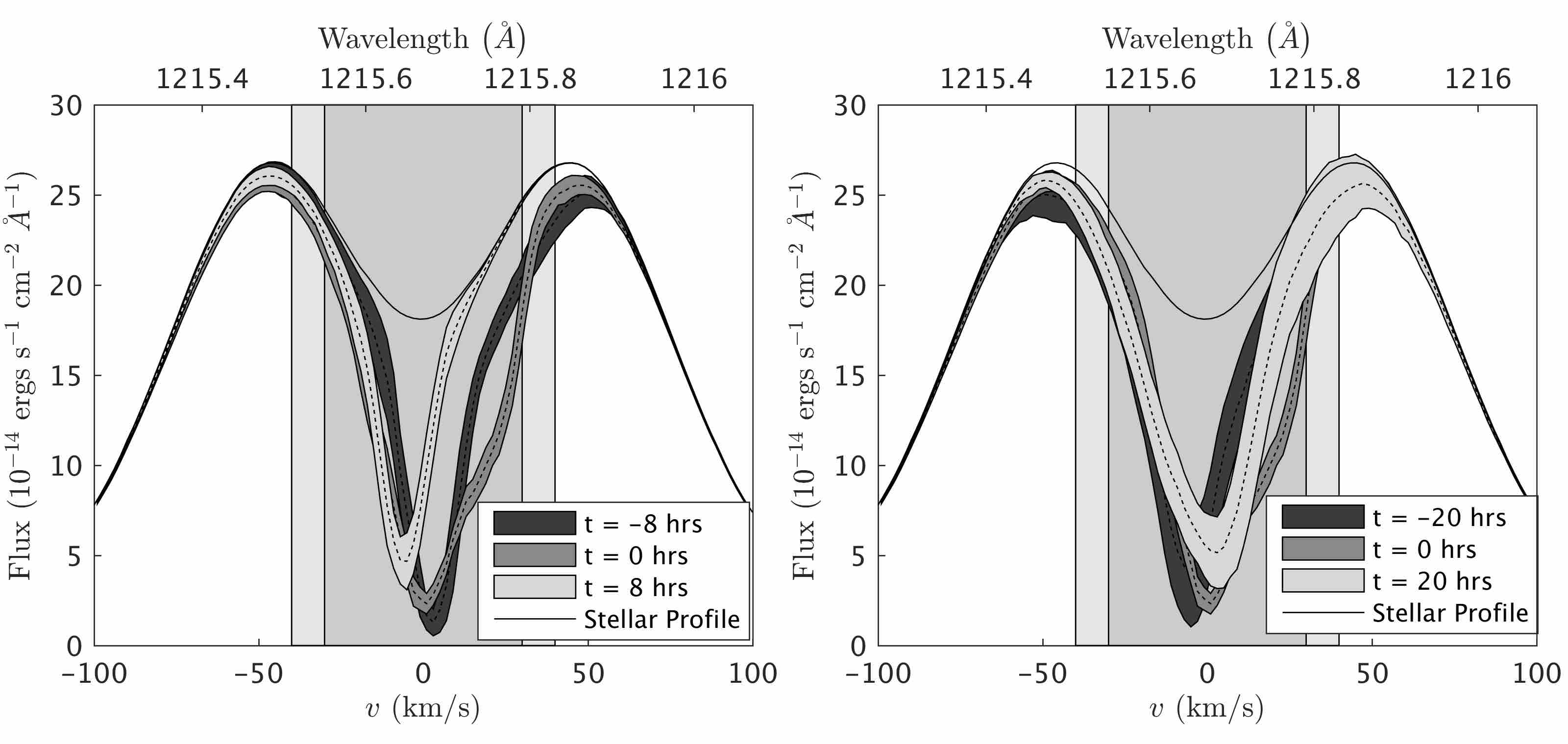}
\caption{Predicted attenuated flux of from star after passing through the planetary wind -  before, during, and after the eclipse.  Thick lines show region within 1 standard deviation of the mean (dashed lines) using data from the last 5 orbits.  Positive velocities are towards the star. Also shown are bands between $\pm 30$ km/s and $\pm 40$ km/s where the interstellar and geo-coronal absorption will likely mask any signal.}
\label{fig:spectra}
\end{figure}

In addition to looking at the normalized intensity, we also generated time averaged theoretical line profiles using the same model for the emission spectra of the H-$\alpha$ line from the star as in \cite{bourrier13}.  In Figure \ref{fig:spectra} the background stellar profile is shown along with the line profile computed from the simulation during the transit as well as $8 \mbox{hrs}$ before and $20 \mbox{hrs}$ after the transit.  The thickness of the lines corresponds to the standard deviation over the last 5 orbits.  

In the left panel of Figure \ref{fig:spectra}, we see the minimum in the line profile (roughly corresponding to peak absorption) shift from positive velocities (at early times) towards negative velocities at later times.  As discussed above, this is due to the Coriolis force bending the radially inward and outward streams up and down orbit respectively.  The absorption profile is also significantly broader during the transit where material leaving the day and night side of the planet still has significant velocities towards and away from the star before being bent by the Coriolis force.  In the right panel of Figure \ref{fig:spectra} we see the opposite trend due to the reversal in the line of sight velocities at around $\pm 60^{\circ}$ corresponding to $\pm 14 \mbox{hrs}$.  We also see increased variability in the line profiles at $\pm 20 \mbox{hrs}$ compared to $\pm 8 \mbox{hrs}$.  Note the line profiles shown ignore interstellar absorption which is expected to wash out most of the line profile between $\pm 30$ or $\pm 40$ km/s.

\section{Radiation pressure}
Radiation pressure was not included in our simulations.  To gain some understanding of its effect we used our semi-analytic models with a range of dimensionless Coriolis radii $\xi_s$ and radiation pressures.  Radiation pressure in the calculations was treated as it is in most simulations in that the gravitational force of the star is reduced by a factor $(1-\alpha)$ where $\alpha$ accounts for absorbtion of momentum in stellar photons.
Thus $\alpha$ is scaled to the gravity of the primary in the limit of $\xi_s \rightarrow 0$.  The ratio of the Coriolis force $2c_s\Omega$ to radiation pressure $\alpha a \Omega^2$ is just $\frac{2 \xi_{\Omega}}{\alpha}$ 

\begin{figure}
\centering
\includegraphics[width=0.95\columnwidth]{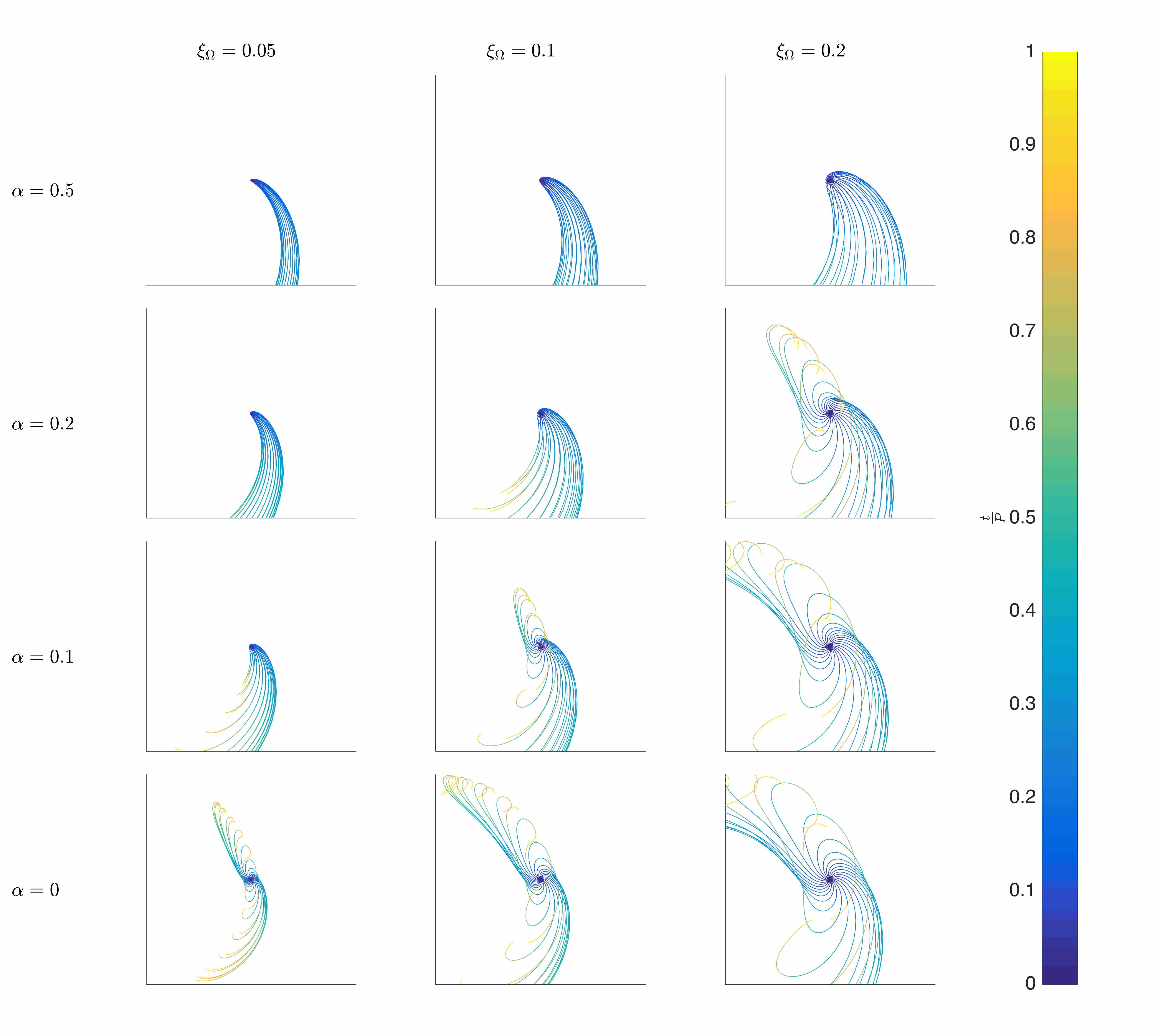}
\caption{Analytic stream lines for various combinations of the dimensionless Coriolis radius $\xi_\Omega$ and dimensionless radiation pressure $\alpha$ (scaled to the stellar gravity).}
\label{fig:streamlines_alpha}
\end{figure}

The trajectories calculated by the semi-analytic method shows that, as expected, the flows can be strongly modified if the outflowing material is universially susceptible to radiation pressure.  As seen in Figure \ref{fig:streamlines_alpha}, the up-orbit stream only appears for $2\xi_{\Omega} \geq \alpha$.  Given the assumption of a constant $\alpha$ everywhere in the flow, in order to have a 'collimated' up-orbit stream the radiative flux must be weak compared to stellar gravity.  This is, of course, in addition to the Coriolis radius being small relative to the orbital separation.  In the next section we discuss interpretation of, and challenges role of including a full treatment of radiation pressure.

\section{Summary of Results and Relation to  Previous Work}

We have carried out high resolution Adaptive Mesh Refinement simulations of the hydrodynamic interaction between a planetary wind and a stellar outflow.  Our simulations included the anisotropic driving of the planetary wind due to day/night temperature differences and we have calculated synthetic transit observations that include the role of ionization of neutral hydrogen in the wind. We have also explored semi-analytic models of wind launching and orbital evolution to understand the role of pressure, tidal and non-interial forces on th flow.

Our results were not directed at any particular exoplanet but highlight general behaviors expected in the global context of massive planets on hot orbits around their host stars.  In this section we review our results in light of previous studies making particular note of open questions raised by our simulations.  We also note that this paper is intended as a first study in a series and  as such, one of its goals was to develop  a simulation platform for studying exoplanet atmospheric blow off as a general phenomena, including construction of a pipeline for synthetic observations. We address the issues raised in the simulations in sequence.
\subsection{Planetary Wind Fall Back}
A complete treatment of the planetary-wind/stellar-outflow interaction problem requires direct calculation of the planetary wind launching by the stellar UV flux.  Detailed modeling of planetary wind launching problem demonstrates that the wind generation falls into two regemes \citep{murrayclay09}. First there is the Energy Limited case where the UV flux energy goes entirely into $pdV$ work launching the  wind.  Second, there is the Radiation-Recombination Limited case in which energy is lost by the flow to recombination and cooling.  Recent studies by \citep{ownalvarez} demonstrate the the details of the launching may admit other possibilities between these two limits. 

In our study, resolution limitations led us to adopt a a boundary condition in which the wind was launched via high temperatures (assumed to be driven by ionization in the atmosphere) at the exobase of the flow.  While we did not explicitly treat the radiation transport within the atmosphere, our model did include a temperature gradient between the day-side (hot) and night-side (cold) as was done in the 2-D models of \citep{stone09}.  Since the night side is too cold to launch a wind our models, we were able to explore the effect of the the wind anisotropy of the global dynamics of the flow.  

As first in explored non-orbital models in \citep{frank15}, the redirection of the strong day-side wind back towards the night side has important consequences including the development of wind material falling back onto the night-side atmosphere.  We found evidence for significant fall back in our models as high as order $\sim 40\%$ of the total outward flow.  This fall back was not seen in previous global models such as \citep{matsakos15} because  an isotropic wind launch boundary condition was used. 2 and 3-D models focusing solely on the planet have, however seen indications of such fallback \citep{tripathi15,christie16}.  It remains to be seen what, if any, the impact of this material raining back on to the planet would have for atmospheric characteristics and dynamics.  The answer will depend on more detailed modeling of how deeply this material can penetrate and how it might change atmospheric radiative transfer and chemical properties.

\subsection{2-Arm Wind Structure}
Our global simulations demonstrated the roll of the coriolis and tidal forces in shaping the planetary wind into upstream and downstream arms. Similar structures were seen in MHD studies of \citep{matsakos15} however those studies did not include the asymmetric launching of the planetary wind.  Using a semi-analytic model we have mapped the trajectory of planetary wind parcels and have demonstrated the dependence of such flow structures on orbital parameters such as $M_p$, $r_p$ and $q$.  We studied how the trajectories depend on the ratio of timescales for orbital motion and wind launching and bending of trajectories by the Coriolis force. We provided an expression for the critical angle $\phi_c$ which the divides the planet into hemispheres which launch an up orbit arm and those which launch a down orbit arm based on the semi-major axis of the wind orbits. We also presented an expression for the "coriolis length" explaining why, in general, only small values of the binary mass ratio and orbital radius will lead to significant departures from a spherical planetary (i.e. secondary) wind. 

Because of the AMR capacities of our code allowed us to track the flow a large distances from the planet at high resolution we saw features such as KH instabilities forming where the planetary wind/stellar outflow interact.  We also see the development of a large stable vortex at in the leeward side of the forward arm. We note that the studies of the global flow pattern which include radiation pressure see only the downstream arm which is interpreted as a cometary tail \citep{schneiter07,cohen09,cohen11,schneiter16}.  As discussed in section 5 (and below) our semi-analytic models also show that for large enough values of the flux (and implied low enough ionization fractions) radiation pressure will remove the forward arm of the flow and push the down-orbit arm to large radii.  We note however that our simulations where meant to map out the general characteristics of winds driven on "hot" orbits.  Thus for stars of lower flux or planetary winds of high ionization the two arm structure we discuss in this study will still occur to some degree.  For example in \citep{bouchy05} consideration was given to the warm Neptune GJ 436 b where radiation pressure was only strong enough decellerate but not halt some part of the forward flow.

\subsection{Synthetic Observations}
Evidence for planetary outflows in the context of Hot Jupiters was first discovered for the star HD 209458.  In particular HST observations implied the existence of material many scale heights above the atmosphere \citep{vidalmadjar03}.  A number of theoretical studies where then able to successfully explain the observations as the result of a planetary wind (e.g. \cite{lammer03,lecavelier04,yelle04,yelle06,baraffe05,tian05,garcia07,schneiter07}).  The focus of this work has been stellar Ly$\alpha$ absorbtion by the outflowing planetary material.  Evidence for both winds and their variability was also seen in the Hot Jupiter orbiting HD189733 \citep{bouchy05}.

In our study we have developed a pipeline for transforming 3-D AMR simulations into synthetic observations of both attenuated stellar line profiles and transits.  For our choice of stellar flux and planetary wind mass loss rates we find that the development of the two armed flow leaves important imprints on the Ly$\alpha$ profile.  In particular we see significant velocity (frequency) asymmetry as up and down orbit material passes in front of the star.  By following the global flow in a large computational domain at relatively high resolution we find that absorbing material can extend around almost the entire orbital domain. In particular we find Coriolis force bends neutral hydrogen leaving the day side of the planet with positive line of sight velocities (i.e. towards the star) up orbit towards positive angles and earlier transit times (Figure \ref{fig:transit}). And likewise, material that flows around the planet due to pressure gradients and heads away from the star (negative line of sight veloci- ties) is bent by the Coriolis force down orbit - towards more negative angles and later transit times. This explains the overall trend towards positive velocities at earlier times and negative velocities at later times. At even earlier times (or more positive angles), we see material that was bent up orbit by the Coriolis force, now being accelerated away from the star due to ram pressure from the stellar wind. This causes the line of sight velocities to decrease as we get further and further up-orbit towards higher angles and earlier times. 

Our synthetic line profiles show shifts from positive velocities (at early times) towards negative velocities at later times (Figure \ref{fig:spectra}). The absorption profile is also significantly broader during the transit where material leaving the day and night side of the planet still has significant velocities towards and away from the star before being bent by the Coriolis force. Finally we note that our profiles show Ly$\alpha$ absorption out to velocities of order $70$ km/s in both blue and redshifted wings.  This is lower than what is observed for systems like HD 209458 and HD189733 and, as such other mechanisms not included in our models (discussed below) are likely at play in those planets.

\subsection{Mechanisms Not Considered}
Our models do not include a number of processes which may be important in some planets experiencing atmospheric blow off.  In particular we have not included radiation pressure, charge exchange or magnetic fields.  In section 5 we used semi-analytic models to explore the role of radiation pressure.  Previous hydrodynamic simulations have demonstrated that radiation pressure on neutral H can drive both remove the up-orbit flow and drive material to velocities to higher velocities as is observed in some systems \citep{schneiter07,cohen09,cohen11,schneiter16}.  Our semianalytic models confirm this result.  It must be noted however that the treatment of radiation pressure in these models utilized a simple prescription in which the gravitational force is uniformly reduced by a factor $1-\alpha$ at every point in the computational domain.  A more detailed treatment would require calculating the radiation transfer and opacity (for a variety of species) through every cell.  Only in that way can the true spatially heterogeneous influence of radiation pressure be determined.  Particle based models including radiation pressure have also been successfully calculated however these do not include the full hydrodynamic behavior of the flow .

Charge exchange has also been show to be an important process for generating absorption at high velocities ($> 100$ km/s:\citep{Holmstrom2008,TremblinChiang2013}.  This occurs as stellar wind protons capture electrons after collisions with lower velocity planetary wind ions. Axi-symmetric hydrodynamic simulations of this process show that depending on the relative strength of the planetary and stellar wind, the resulting fast H atoms will be either be confined to a sheath around the now shock or can penetrate the subsonic regions of the wind. In general the population of fast H atoms is not expected to alter the global flow dynamics \citep{christie16}.

Finally magnetic fields, if present, can be expected to play an important role in both the launching of the planetary wind and its subsequent interaction with the stellar outflow. As has been shown by a number of authors \citep{Trammellea2011,Trammellea2014,Adams2011,OwenAdams2014} planetary fields can alter extent and geometry of planetary mass loss.  At larger distances interaction between the planetary and stellar field can control subsequent flow dynamics. In \cite{matsakos15} a classification system for the flows was developed based on the relative strength of both planetary and stellar flows and fields.  Of particular interest is the ability of fields to channel planetary wind material onto the star leading, perhaps, to enhanced X-ray emission \citep{Shkolnikea2008, Pillitteriea2014}.  In the present study we focused on hydrodynamic interactions to establish a baseline in accounting for the mechanisms effecting global flow patterns. The next study in this series will include the effect of magnetic fields.

We also note that future work can include a better treatment of heating and cooling processes in the global flow \citep{schneiter16}. The assumption of tidal locking for the planet should also be relaxed allowing modeling of wind launching from faster rotating planets or, along a similar vein, a treatment of launching from planets with strong atmospheric circulation \citep{tripathi15,Dobbs-DixonAgol2013}

\section{Acknowledgments}
We thank James Kasting for many discussions related to planetary winds. This work used the computational and visualization resources in the Center for Integrated Research Computing (CIRC) at the University of Rochester. Part of the work also used the Extreme Science and Engineering Discovery Environment (XSEDE), which is supported by National Science Foundation grant OCI-1053575. Financial Support for this project was provided by the Department of Energy grant GR523126, the National Science Foundation grant GR506177, and the Space Telescope Science Institute grant GR528562.

\appendix
\section{Local accelerations around Planet}
\label{app:localforces}
We begin with the acceleration felt by a particle in the co-rotating frame located at position $\mathbf{r}$ traveling at velocity $\mathbf{v}$
\begin{equation}
  -\frac{GM_{\star}}{|r_{\star}|^3} \mathbf{r_{\star}} - \frac{GM_p}{|r_p|^3}\mathbf{r_p} - 2 \mathbf{\Omega} \times \mathbf{v} - \mathbf{\Omega} \times \left ( \mathbf{\Omega} \times \mathbf{r} \right )
\end{equation}
where $\mathbf{r_{\star}}$ and $\mathbf{r_p}$ are the positions of the particle relative to the star and planet respectively, $M_\star$ and $M_p$ are the masses of the star and planet respectively, $\mathbf{\Omega}$ is the angular velocity of the corotating frame and $\mathbf{a}$ is the location of the planet relative to the star.  We can expand this force about the planet using
$\mathbf{r_{\star}} = \mathbf{a}+\mathbf{r_p}$ and $\mathbf{r} = \mathbf{a} \left ( 1 - \frac{1}{q+1} \right ) + \mathbf{r_p}$.  This gives

\begin{equation}
\begin{aligned}
    & -\frac{GM_{\star} \left ( \mathbf{a} + \mathbf{r_p} \right)}{\left ( a^2+2 \mathbf{a} \cdot \mathbf{r_p} + r_p^2 \right )^{3/2}} 
    - \frac{GM_p}{|r_p|^3}\mathbf{r_p} \\
    & - 2 \mathbf{\Omega} \times \mathbf{v}  
    + \Omega^2 \left ( \mathbf{a} - \frac{q}{q+1}\mathbf{a} + \mathbf{r_p} - \left ( \mathbf{r}\cdot \mathbf{\hat{\Omega}} \right ) \mathbf{\hat{\Omega}} \right )..
\end{aligned}
\end{equation}

We restrict ourselves to the orbital plane $( \mathbf{r} \cdot \mathbf{\hat{\Omega}} = 0)$ and expand to first order in $\frac{r_p}{a}$ and $q$ and use $\Omega^2 = \frac{G \left (M_{\star}+M_p \right)}{a^3}$ to get
\begin{equation}
\begin{aligned}
    & -\frac{GM_{\star} \left ( \mathbf{a} + \mathbf{r_p} \right) \left ( 1-3 \frac{\mathbf{a} \cdot \mathbf{r_p}}{a^2} \right )}{a^3}   
    - \frac{GM_p}{|r_p|^3}\mathbf{r_p} \\
    & - 2 \mathbf{\Omega} \times \mathbf{v}  
    + \frac{G \left ( M_{\star} + M_p \right)}{a^3} \left ( \mathbf{a} - \frac{q}{q+1}\mathbf{a} + \mathbf{r_p} \right ).
\end{aligned}
\end{equation}
This can be written as
\begin{equation}
    3 \Omega^2 \mathbf{\hat{a}} \cdot \mathbf{r_p} \mathbf{\hat{a}}  - \frac{GM_p}{|r_p|^3}\mathbf{r_p} - 2 \mathbf{\Omega} \times \mathbf{v}  
\end{equation}
Now the difference between the pressure from the Parker wind and gravitational attraction of the planet is what provides the acceleration in the Parker wind.  This acceleration is just 
\begin{equation}
    \frac{d v}{dt} = \frac{dv}{dx}\frac{dx}{dt}=\frac{dv}{dx}v = \frac{1}{2}\frac{dv^2}{dx} = \frac{c_s^2}{2 r_s} \frac{d \psi}{d \Xi}
\end{equation}
Including the Parker pressure gives us
\begin{equation}
   3 \Omega^2 \mathbf{\hat{a}} \cdot \mathbf{r_p} \mathbf{\hat{a}} + \frac{c_s^2}{r_s}\frac{d \psi}{d \Xi} \mathbf{r_p} - 2 \mathbf{\Omega} \times \mathbf{v}  
\end{equation}

\section{Orbital trajectories}
\label{app:orbitaltrajectories}
  In the limit of $q << 1$,  particles leaving the sonic surface at $r_s$ at velocity $c_s$ will have initial position $(\mathbf{r'})$ and velocity $(\mathbf{v'})$ in the inertial frame
\begin{equation}
 \begin{aligned}
 \mathbf{r'}= &\mathbf{a} + r_s \mathbf{\hat{n}} = a \left ( \mathbf{\hat{a}} + \xi_s \mathbf{\hat{n}} \right ) \\
  \mathbf{v'} = & \mathbf{\Omega} \times \mathbf{r'} + c_s \mathbf{\hat{n}} \\
   = & a \Omega \left (\mathbf{\hat{\Omega}} \times \mathbf{\hat{a}}+\xi_s \mathbf{\hat{\Omega} \times \hat{n}} + \xi_\Omega \mathbf{\hat{n}} \right )
\end{aligned}
\end{equation}
where $\Omega$ is the orbital frequency, $\mathbf{a}$ is the position of the planet relative to the stellar companion, $\mathbf{\hat{n}}$ is the normal vector at the location on the surface, $\xi_{\Omega} = \frac{c_s}{a\Omega}$ and $\xi_s = \frac{r_s}{a}$.

Now the specific energy of each orbit as 
\begin{equation}
    \epsilon' = \frac{1}{2}v'^2-\frac{GM}{r'}=\epsilon \left (\frac{2a}{r'} - \frac{v'^2 }{a^2 \Omega^2}  \right ),
\end{equation}  
where we have expressed it in terms of the specific energy of the planetary orbit $\epsilon=-\frac{GM}{2a}=-\frac{a^2\Omega^2}{2}$.

Now to first order in $\xi_\Omega = \frac{c_s}{v_k}$ and $\xi_s = \frac{r_s}{a}$ we have

\begin{equation}
    \frac{a}{r'} = \frac{1}{1+\xi_s \mathbf{\hat{n}} \cdot \mathbf{\hat{a}}} \approx 1 - \xi_s \mathbf{\hat{n}} \cdot \mathbf{\hat{a}}
\end{equation}
and
\begin{equation}
    \frac{v'^2}{v_k^2} = 1 + 2\xi_s \mathbf{\hat{n}} \cdot \mathbf{\hat{a}} + 2 \xi_\Omega \mathbf{\hat{n}} \cdot \left ( \mathbf{\hat{\Omega}} \times \mathbf{\hat{a}} \right )
\end{equation}

Combining these gives
\begin{equation}
    \frac{\epsilon'}{\epsilon}=1 - 2 \xi_\Omega \sin{\phi} \sin{\theta} - 4 \xi_s \cos{\phi}\sin{\theta}
\end{equation}
where $\phi$ is the angle measured from $\mathbf{\hat{a}}$ around $\mathbf{\hat{\Omega}}$ and $\theta$ is the angle measured down from the axis of rotation $\mathbf{\hat{\Omega}}$.

We can also rewrite this as 
\begin{equation}
    \frac{\epsilon'}{\epsilon}=1 - 2\xi_T  \sin \theta \sin{ \left ( \phi-\phi_c \right ) },
\end{equation}
where $\phi_c = \arctan \frac{-2 \xi_s}{\xi_\Omega}$ with $-\pi/2 < \phi_c < 0$ and $\xi_T = \left ( \xi_\Omega^2 + 4 \xi_s ^2 \right )^{1/2}$.

Since $a'=-\frac{GM}{2\epsilon'}$ we also have for the semi-major axis of each orbit
\begin{equation}
    \frac{a'}{a}=1 + 2\xi_T  \sin \theta \sin{ \left ( \phi-\phi_c \right ) }
\end{equation}
Material leaving from $\phi_c < \phi < \phi_c+\pi$ will have a larger semi-major axis and a longer period and will form the down-orbit stream, while material leaving from $\phi_c + \pi < \phi < \phi_c + 2 \pi$ will have a smaller semi-major axis and a shorter period and will form the up-orbit stream.

Now we can take the surface average of the specific orbital energy  over the two hemispheres where $+$ and $-$ corresponds to the change in semi-major axis and refer to the down-orbit and up-orbit streams respectively.  

\begin{equation}
    \frac{\bar{\epsilon'}_\pm}{\epsilon}= 1\mp\frac{4}{\pi}\xi_\Omega 
\end{equation}
which then gives us an 'average' semi-major axis for the streams of
\begin{equation}
      \frac{\bar{a'}_\pm}{a}=  1\pm \frac{4}{\pi}\xi_\Omega
\end{equation}
Now since $\Omega \propto a^{-3/2}$ so we can calculate the change in angular velocity 
\begin{equation}
    \frac{\Omega'_\pm}{\Omega}=1 \mp  \frac{6}{\pi} \xi_\Omega
\end{equation}
and the angular distance the streams will travel down or up orbit per orbital period in the co-rotating frame 
\begin{equation}
    \Theta_D\pm = 2 \pi \left ( \Omega'_\pm-\Omega \right ) = \mp 12 \xi_\Omega
 \end{equation}

\bibliography{hotJupitor.bib}

\bsp

\label{lastpage}

\end{document}